\documentclass[apj]{emulateapj}
\usepackage{lscape}
\usepackage{amsmath}
\usepackage{graphicx}
\usepackage{natbib}
\usepackage{longtable}

\newcommand{\hbeta}{H{$\beta$}}

\newcommand{\CIV}{C{\sevenrm IV}}

\newcommand{\MgII}{Mg{\sevenrm II}}

 \font\sevenrm=cmr7 scaled 1000

\begin{document}

%\title{The Luminosity Function and Black Hole Mass Function of Broad Line Quasars}
\title{The Demographics of Broad Line Quasars in the Mass-Luminosity
  Plane. II. \\
  Black Hole Mass and Eddington Ratio Functions}

\shorttitle{QUASAR DEMOGRAPHICS II}

%\slugcomment{Draft Version}

\shortauthors{KELLY \& SHEN}
\author{Brandon C. Kelly\altaffilmark{1} and Yue Shen\altaffilmark{2}}
%Harvard-Smithsonian Center for Astrophysics, 60
%Garden Street, Cambridge, MA 02138, USA.}
\altaffiltext{1}{Department of Physics, Broida Hall, University of
  California, Santa Barbara, CA, 93107}
\altaffiltext{2}{Harvard-Smithsonian Center for Astrophysics, 60
Garden Street, MS-51, Cambridge, MA 02138, USA.}

\begin{abstract}
We employ a flexible Bayesian technique to estimate the black hole mass and
Eddington ratio functions for Type 1 (i.e., broad line) quasars from a
uniformly-selected data set of $\sim 58,000$ quasars from the SDSS DR7. We
find that the SDSS becomes significantly incomplete at $M_{BH} \lesssim 3
\times 10^8 M_{\odot}$ or $L / L_{Edd} \lesssim 0.07$, and that the number
densities of Type 1 quasars continue to increase down to these limits. Both
the mass and Eddington ratio functions show evidence of downsizing, with the
most massive and highest Eddington ratio black holes experiencing Type 1
quasar phases first, although the Eddington ratio number densities are flat
at $z < 2$. We estimate the maximum Eddington ratio of Type 1
quasars in the observable Universe to be $L / L_{Edd} \sim 3$. Consistent with our results in Paper I,
we do not find statistical evidence for a so-called ``sub-Eddington
boundary'' in the mass-luminosity plane of broad line quasars, and
demonstrate that such an apparent boundary in the observed distribution can
be caused by selection effect and errors in virial BH mass estimates. Based
on the typical Eddington ratio in a given mass bin, we estimate growth
times for the black holes in Type 1 quasars and find that they are 
comparable to or longer than the age of the universe, implying an
earlier phase of accelerated (i.e., with higher Eddington ratios) and
possibly obscured growth. The large masses probed by our sample imply that
most of our black holes reside in what are locally early type galaxies, and
we interpret our results within the context of models of self-regulated black
hole growth.
\end{abstract}

\section{Introduction}\label{sec:intro}

\subsection{Background and Motivation}

Understanding how and when supermassive black holes (SMBHs) grow is currently
one of the great outstanding problems in extragalactic astronomy. The
evolution of the galaxy and SMBH populations are not independent, as implied
by established correlations between the mass of the SMBH, $M_{BH}$, and
properties of the host galaxy bulge, such as luminosity
\citep{korm95,mclure01,mclure02}, stellar velocity dispersion \citep[the
$M_{BH}$--$\sigma_*$ relationship, e.g.,][]{gebh00a, merr01, trem02},
concentration or Sersic index \citep{graham01,graham_driver07a}, bulge mass
\citep{mag98,marc03,haring04}, and binding energy \citep{aller07,fundplane}.
Motivated by these empirical trends, a number of authors have invoked
AGN\footnote[3]{In this work we will use the terms AGN and quasar to refer to
the same type of object. No luminosity difference is implied except
where explicitly stated.} feedback as a
means of regulating the SMBH's growth, which ties $M_{BH}$ to properties of
the host galaxy
\citep[e.g.,][]{silk98,fabian99,begel05,murray05,dimatt05,springel2005,hopkins_long,johan09}.
In addition to regulating the growth of SMBHs, AGN feedback has been invoked
as a means of quenching the growth of the most massive galaxies
\citep[e.g.,][]{bower06,croton06}. Alternatively, it has been suggested that
the origin of the scaling relationships emerges from the stochastic nature of
the hierarchical assembly of black hole and stellar mass through galaxy
mergers \citep{peng07,jahnke11}. In addition, more recent observational results
have painted a more complicated picture, and it is unclear to what degree the scaling
relationships between the host galaxy and $M_{BH}$ extend beyond
classical bulges or down to lower black hole masses
\citep[e.g.,][]{hu08,graham08,gult09,greene2010a,jiang11,korm11,graham2012}.

The galaxy and SMBH populations are also coupled because the fueling of
SMBHs, which initiates AGN activity, depends on events that occur within and
to the host galaxy. Many models have invoked major mergers of two gas-rich
galaxies as the triggering mechanism for quasar activity that
provides the bulk of SMBH growth
\citep[e.g.,][]{sanders1988,Sanders_Mirabel_1996,kauffmann2000,Wyithe_Loeb_2003,springel2005,dimatt05,sijacki07,dimatteo08,hopkins2008,somer08,shen09}.
Models have also invoked large-scale secular instabilities as a means of
fueling AGN activity and growing SMBHs in disks
\citep[e.g.,][]{bower06,bournaud2011,fanidakis2011}. At lower luminosities or
redshifts, other fueling mechanisms may dominate the triggering of AGN
activity \citep[e.g.,][]{hh09}, such as external interactions
\citep[e.g.,][]{serber06,alonso07,woods07,silverman2011,ellison2011,liu2012}, stochastic accretion of
gas \citep[e.g.,][]{stochacc}, bar instabilities
\citep[e.g.,][]{shlosman89,garcia05,hq2010}, or stellar mass loss
\citep[e.g.,][]{norman88,ciotti97,ciotti07,kauffmann2009,ho2009}. Regardless
of the fueling mechanism, if the SMBH's growth is self-regulated via AGN
feedback then the final mass of the SMBH after a fueling event is set by the
binding energy of the bulge \citep{younger08}.

% While it is established that the galaxy and SMBHs populations are not
% completely independent, recent observational work has painted a more
% complicated picture regarding the connection between
% these two populations. In general, it is unclear if the scaling
% relationships between the host galaxy and $M_{BH}$ extend beyond
% classical bulges. The scatter in the $M_{BH}$--$\sigma_*$ relation is
% larger for spirals \citep{graham08,gult09}, and appears to increase
% at low $M_{BH}$ such that most SMBHs lie below the
% $M_{BH}$--$\sigma_*$ relation, \citep[e.g.,][]{greene2010a}. Several authors
% have found evidence that there is no significant correlation between
% $M_{BH}$ and pseudobulge luminosity
% \citep{hu08,greene08,graham08,graham09,jiang11,korm11}. Recently, \citet{korm11}
% argue that $M_{BH}$ also does not correlate with properties of galaxy
% disks, while \citet{graham2011} concluded that $M_{BH}$ does correlate
% with $\sigma_*$ for barred galaxies, even though
% their $M_{BH}$--$\sigma_*$ relationship is offset
% from that of non-barred galaxies. These studies imply that the fueling
% mechansims of AGN activity, as well as the role of feedback or other
% mechanisms for regulating the mass of the SMBH, are diverse and not
% necessarily uniform for all galaxy types.

Ideally, in order to constrain models of SMBH fueling, growth, and impact on
the host galaxy one would like to follow across time the stochastic process
that is the evolution of SMBHs and their hosts. However, this is not
possible, so demographic studies must be used to reconstruct the evolution of
these populations. Because of this, the demographics of SMBHs, AGN, and their
host galaxies is one of the primary empirical tools that we have for placing
constraints on astrophysical models of SMBH growth and fueling.
Traditionally, studies of AGN demographics have focused on the luminosity
function \citep[e.g., for some recent estimates
see][]{boyle2000,fan2001,ueda2003,wolf2003,fan2004,barger2005,hasinger2005,richards2005,richards2006,fontanot2007,bongiorno2007,hopkins_lumfunc,silverman2008,croom2009,jiang2009,aird2010,willott2010b,fiore2012,shen2012}.
Many recent luminosity function studies lead to the important discovery that the comoving number
densities of more luminous AGN peak at earlier times than do the number
densities of less luminous AGN
\citep[e.g.,][]{cowie2003,steffen2003,ueda2003,hasinger2005,bongiorno2007,croom2009,rigby2011},
a phenomenon termed `cosmic downsizing'. That the more luminous AGN
population turns-off at earlier cosmic epochs implies that the more massive
SMBHs grow first.

The luminosity function is the convolution of the black hole mass function
with the Eddington ratio distribution, weighted by the duty cycle of
AGN activity; the duty cycle is the fraction of SMBHs that are `active'
at any time. Thus, while easy to measure, the luminosity function
provides somewhat limited physical insight. Fortunately, we have other
empirical tools based on AGN demograhics which provide various
complementary observational constraints on SMBH growth and
fueling. These include spatial clustering of AGN, AGN host galaxy
properties, and black hole mass and Eddington ratio distributions as a
function of redshift. Spatial clustering provides observational
constraints on the masses of AGN host dark matter halos, the bias of
AGN environments, and AGN duty cycle. AGN host galaxy properties, on
the other hand, provide insight into the fueling and triggering of AGN
activity, as well as the effects of AGN feedback. And finally, studies
of SMBH mass and Eddington ratio distributions provide insight into
the growth of supermassive black holes and AGN duty cycles. These
various empirical tools provide different but complementary
information, and it is the goal of this work to further improve our
understanding of SMBH mass and Eddington ratio distributions for Type
1 quasars.

% Studies of AGN host galaxy demographics have also found evidence that
% star-formation rate and SMBH accretion rate, as traced through the AGN
% luminosity, are correlated both on a cosmological level
% \citep[e.g.,][]{silverman2008,aird2010} and within individual galaxies
% \citep[e.g.,][]{imanishi2004,schweitzer2006,shi2007,netzer2007b,lutz2008,shi2009,mor2012}.
% These results imply that SMBH fueling and growth is at least in part related
% to the presence of star formation, which is perhaps not surprising as both
% can be initiated by the presence of cold gas. Recently, \citet{diamond2012}
% found that in local Seyfert galaxies the correlation between star formation
% rate and SMBH accretion rate is significantly stronger when limiting the
% analysis to the nuclear star formation rate, in agreement with results on
% SMBH fueling obtained from smoothed particle hydrodynamic simulations
% \citep{hq2010}.

The black hole mass function (BHMF) and black hole Eddington Ratio
function (BHERF) quantify the comoving
number density of SMBHs as a function of $M_{BH}$ and $L /
L_{Edd}$. Therefore, the BHMF and BHERF provide a complete census of
the SMBH population with respect to $M_{BH}$ and $L / L_{Edd}$,
providing important information for constraining models of
SMBH growth. Indeed, many models for SMBH growth have made predictions
for the BHMF and BHERF at a variety of redshifts
\citep[e.g.,][]{catt05,dimatteo08,hopkins2008,shen09,tan09,vol10,fanidakis2011,fanidakis2012,natarajan2012,draper2012}. For a review of the BHMF of SMBHs, see \citet{bhmf_review}.

There are two common approaches for estimating the BHMF and BHERF. The
first approach is to employ a continuity equation describing the
evolution of the SMBH population, and has its roots in the work of
\citet{soltan82}. The basic idea behind these continuity-equation
methods is to assume a BHMF at very high redshift, use the AGN luminosity function as a
probe of how fast the BHMF is changing, and then evolve the BHMF using
the local BHMF as a constraint. The continuity equation methods
also provide an estimate of the typical duty cycle and radiative
efficiency of SMBHs, which can then be linked to a typical quasar
lifetime and black hole spin, respectively. Several
authors have derived the local BHMF for all SMBHs from individual sources using
the scaling relationships between $M_{BH}$ and host spheroidal properties
\citep[e.g.,][]{sal99,yu02,aller02,marc04,shank04,tundo07,yu08,shank09,vika09},
providing the needed integral constraint for the continuity equation method. A
number of groups have employed
variations on the continuity equation technique
\citep[e.g.,][]{yu02,marc04,merloni04,hopkins_lumfunc,merloni08,shank09,cao10},
and have generally concluded that most, if not all, of the local BHMF can be
explained as the relic of AGN activity, with SMBH growth being dominated by
periods when the SMBH was radiating near the Eddington limit. In addition,
these studies have generally concluded that SMBH growth is anti-hierarchical,
in agreement with the cosmic downsizing seen in the AGN luminosity function
studies.

The other commonly employed approach to estimating the BHMF and BHERF is to use
scaling relationships to obtain estimates of $M_{BH}$ for individual
sources, and then derive the BHMF and BHERF from the distribution of
these estimates. Indeed, this approach provides the local BHMF needed
for the continuity equation approach. It is also possible to
incorporate scaling relationships into the continuity equation methods,
as done by \citet{merloni08}. The advantage of the scaling
relationship approach is that the BHMF and BHERF at a given redshift
is derived from the distribution of $M_{BH}$ estimates, providing more
information at that redshift than is provided by the local BHMF integral constraint used in the
continuity equation methods. Unfortunately, the scaling
relationship methods have the disadvantage that the
host galaxy scaling relationships are currently only used to estimate the $z \approx 0$
BHMF and BHERF for all SMBHs. Many studies have found evidence for
evolution in the SMBH-host galaxy scaling relationships 
\citep[e.g.,][but see Lauer et al. 2007, Shen \& Kelly 2010, Schulze
\& Wisotzki 2011, and Portinari et al. 2012 for cautionary
notes]{treu04,peng06,treu07,woo08,decarli10,merloni10,bennert10}, and,
while there have been attempts to estimate the BHMF based on an assumed form
of the evolution in the scaling relationships \citep{tam2006,sbh09,li2011},
the quantitative form of the evolution has not been sufficiently precise to
motivate widespread use of the scaling relationships to estimate BHMFs
beyond the local universe.

For Type 1 quasars, an alternative to employing the host galaxy
scaling relationships to estimate $M_{BH}$ is to employ so-called
virial mass estimates
\citep[e.g.,][]{wandel1999,mclure_jarvis2002,vest02,vest06}. These
virial mass estimates are derived by using the width of the broad
emission lines as a proxy for the velocity dispersion of the clouds
emitting the broad emission lines, and the luminosity as a proxy for
the broad line region size
\citep{kaspi2005,bentz2009a,greene2010b}. The virial mass estimates
have a statistical scatter about the mass estimates derived from
reverberation mapping \citep{peter2004,bentz2009b} of $\sim 0.4$ dex
\citep[e.g.,][]{vest06,park2012b}, although there may be additional
systematic errors
\citep{krolik01,collin2006,shen08,fine08,marconi08,denney09,rafiee2011a,steinhardt2011}.
The virial and reverberation mapping mass estimates are calibrated to
the local $M_{BH}$--$\sigma^*$ relationship
\citep{onken2004,woo2010,graham2011,park2012a}.

The virial mass estimates have the advantage that they can be used
beyond the local universe, providing estimates of the BHMF and BHERF
for active SMBHs directly from estimates of $M_{BH}$. However, they
have the disadvantage that they only estimate the BHMF for Type 1
quasars, which are a subset of the active SMBH population, which in
turn is a subset of the entire SMBH population. However, despite this,
the BHMF and BHERF for Type 1 quasars still provide an important
observational constraint on models for SMBH growth that is
complementary to other empirical tools based on AGN
demographics. Compared to luminosity, studies of AGN demographics with
respect to $M_{BH}$ have the advantage that mass is a more stable
quantity, in that $M_{BH}$ can only increase and does so during active
phases or through SMBH mergers. As a result, demographics of AGN
$M_{BH}$ probe the subset of SMBHs that are actively growing at any
given $z$ and $M_{BH}$. This implies that, for example, the Type 1
quasar BHMF provides an estimate of the SMBH duty cycle for Type 1
quasar activity as a function of $M_{BH}$ and $z$, provided one has an
estimate of the BHMF for all SMBHs at that redshift. Moreover,
estimates of SMBH growth times for Type 1 quasars, calculated from the
BHMF and BHERF, can also provide constraints on models for SMBH
growth, especially at high redshift. In particular, the discovery of
Type 1 quasars with $M_{BH} \sim 10^9 M_{\odot}$ out to $z \sim
6$--$7$ \citep{jiang2007,kurk2007,willott2010a,mortlock11} places
strong constraints on models for the formation and growth of SMBH
seeds \citep[e.g.,][]{haiman01}.

Numerous recent studies have used the virial mass estimates to study
the demographics of Type 1 quasar $M_{BH}$, Eddington ratio, and their
evolution
\citep[e.g.,][]{Woo_Urry_2002,vest04,mclure04,koll06,sulentic2006,babic2007,netzer07a,shen08,gavignaud2008,fine08,trump2009,trump2011,trakh2011,rafiee2011b}.
In particular, the use of virial mass estimates has led several
authors to estimate the black hole mass function BHMF and BHERF for Type 1 quasars,
\citep{wang2006,greene2007,vest08,vest09,kelly2009,schulze2010,kelly2010,willott2010a,shen_kelly2010,shen2012,
nobuta2012}.  In addition, $M_{BH}$ is a
fundamental physical parameter of black hole accretion flows, making
the BHMF and BHERF important for studies of accretion physics, as it
describes which regions of the $M_{BH}$--$L / L_{Edd}$ plane are
probed by current and future surveys. Based on AGN number densities,
several groups have found evidence for downsizing in $M_{BH}$ of SMBHs
in Type 1 quasars
\citep[e.g.,][]{vest09,labita2009a,labita2009b,kelly2010,shen2012},
implying that at least some of the downsizing in the AGN luminosity
function may be driven by downsizing in $M_{BH}$. In addition, the
Eddington ratio distributions derived by 
\citet[][hereafter K10]{kelly2010} and \citet{shen2012} imply that at $z > 0.5$ there is a broad range in $L / L_{Edd}$
for Type 1 quasars, and that most
Type 1 quasars are not radiating near the Eddington limit; similar
results were obtained by \citet{schulze2010} at $z < 0.3$.

This is the second paper in a two-part series to study the demographics of
Type 1 quasars out to $z \sim 4.75$ in the two-dimensional quasar mass and
luminosity space. Such studies represent a next step in Type 1 quasar
demographic studies by placing joint constraints on the distributions of
their instantaneous activity (i.e., luminosity function) and assembly history
(i.e., BH mass function) via the Eddington ratio distribution. As such, the
2D demographic studies provide an advanced view of the cosmic growth of the
SMBH population compared to using the luminosity function or BHMF alone. Our
study is motivated by both the recent availability of $\sim 58,000$
uniformly selected Type 1 quasars from the SDSS DR7 with virial mass
estimates from \citet{shen2011}, as well as the recent advancements in
statistical techniques for studying AGN demographics. The incredible size of
this data set combined with our new powerful statistical tools provides us
with an unprecedented ability to characterize the demographics of Type 1
quasars at a variety of redshifts. The broad redshift range of our sample
($0.4 < z < 4.75$) allows us to probe the evolution of a subset of the
actively growing SMBH population during the epochs over which $\sim 95\%$ of
SMBH growth occured \citep[e.g.,][]{shank09}. 

\subsection{Comparison with \citet{shen2012}}

In the first paper of this
series, \citet[][hereafter Paper I]{shen2012}, we presented our sample as
well as the binned estimates of the luminosity function and the BHMF. We also
extended the statistical method of K10 to estimate the BHMF independently in
different redshift bins, and to include a luminosity-dependent bias term for
the virial mass estimates. The former improvement enables us to study the
evolution of the BHMF and BHERF without assuming a parameteric form for this
evolution, while the latter improvement enables us to study the effects of a
systematic error in the virial mass estimates with luminosity. A
luminosity-dependent bias term is motivated by the small scatter in the
virial mass estimates at fixed $M_{BH}$ and $L$, as observed by several
recent studies of AGN demographics
\citep[][K10]{koll06,shen08,fine08,steinhardt2010b,shen_kelly2010}. The
results from our model implied that virial mass estimates derived from the
FWHM of the \MgII\ and \CIV\ emission lines systematically overestimate
$M_{BH}$ at higher than average luminosity, assuming that the statistical
scatter in the mass estimates at fixed $M_{BH}$ and $L$ is constant.

In Paper I we also used our model to study the luminosity function of Type 1
quasars, and presented the BHMF and BHERF derived from our model. We
studied the luminosity function below the flux limit, assuming our model for
the SMBH mass and Eddington ratio distributions. We found evidence for
downsizing in both the luminosity function and BHMF. In addition, we found
that our model luminosity function makes reasonable predictions when
extrapolated to $\sim 3$ mag fainter than our SDSS DR7 sample, and the
extrapolations were often consistent with luminosity functions estimated from
deeper surveys. We also found evidence that the average Eddington
ratio of Type 1 quasars increases towards higher redshift, in qualitative
agreement with the conclusions of continuity equation methods
\citep[e.g.,][]{shankar2012} and predictions from recent hydrodynamic simulations of SMHB
growth at $z>4$ \citep[e.g.,][]{DeGraf_etal_2012}.

In this paper we build upon the work of Paper I and focus our analysis and
discussion on the BHMF, the BHERF, and quantities that can be derived from
them. The main differences between Paper I and this paper are:
\begin{itemize}
  \item 
    We make several improvements to our statistical model. We extend
    the statistical model of K10 and Paper I to model the joint
    distribution of Eddington ratio and $M_{BH}$ at a given redshift as a
    mixture of 2-dimensional log-normal distributions. This provides us
    more flexibility compared to Paper I,
    where we assumed that the Eddington ratio distribution at fixed $z$
    and $M_{BH}$ could be described as a single log-normal distribution
    with geometric mean depending linearly on $\log M_{BH}$. In addition,
    we also use a Student's $t$-distribution to model the distribution of
    measurement errors in the mass estimates, which downweights outliers
    compared to the previously-used Gaussian distribution. This latter improvement
    was incorporated to make our results more robust against outliers in
    the mass-luminosity plane, as such sources may be subject to
    unidentified systematic error. We made these improvements to test the robustness of the key conclusions from Paper I.
    \item
      Because we have employed a more flexible model for the BHERF,
      the focus of this paper is on the BHMF and BHERF of Type 1
      quasars. As such, in this paper we discuss
      the BHMF and BHERF in greater depth than we did in Paper I, and discuss the
      distribution of Eddington ratios at fixed $M_{BH}$. In addition,
      we discuss what the derived BHMF and BHERF imply with regard to
      the growth of supermassive black holes. 
    \item
      We focus our presentation and discussion on the BHMF and BHERF
      based on the usual assumption that the mass estimates are
      unbiased. This is the assumption that is typically made in the
      literature, but is different from Paper I where we allowed the
      mass estimates to have a luminosity-dependent bias. In this paper we
      present the results for the model assuming the estimates are
      unbiased, but also discuss how allowing a luminosity-dependent
      bias changes the results. Between Paper I and this paper we
      cover both simple models for the behavior of the virial mass
      estimates with respect to the BHMF and BHERF. 
    \item
      Unlike in Paper I, we do not present or discuss the optical
      luminosity function implied by our model, as there is little
      difference from Paper I, nor is the luminosity function the
        focus of this paper. Instead, we focus on the BHMF and
      BHERF. 
      \item
        Paper I goes into greater depth with respect to biases in the
        virial mass estimates, and what we can conclude about such
        biases based on AGN demographics. That is not the focus of
        this paper. 
\end{itemize}
Between Paper I and this paper we present an in-depth analysis of
Type 1 quasar demographics with respect to the properties of their
virial mass estimates, their luminosity function, their black hole
mass function, their black hole Eddington ratio function, and their
behavior in the mass-luminosity plane. 

This paper is organized as follows. We describe our statistical model in
\S\ref{s-statmodel}. We present our estimated BHMF in \S\ref{s-bhmf} and
BHERF in \S\ref{s-bherf}, and discuss our results in \S\ref{s-discussion}. We
summarize our results in \S\ref{s-summary}. Throughout the paper we adopt a
flat $\Lambda$CDM cosmology with cosmological parameters
$\Omega_{\Lambda}=0.7$, $\Omega_0=0.3$, $h=0.7$, to match most of the recent
quasar demographics studies. Volume is in comoving units unless otherwise
stated. We distinguish virial masses from true masses with a subscript $_{\rm
vir}$. For simplicity, quasar luminosity is expressed in terms of the
rest-frame 2500\,\AA\ continuum luminosity ($L\equiv \lambda L_\lambda$). The
conversion between the 2500\,\AA\ continuum luminosity and the absolute
$i$-band magnitude normalized at $z=2$ ($M_{i}[z=2])$ is given by eqn. (4) in
\citet{richards2006}. Lower case letters refer to logarithms of quantities
based on mass or luminosity, e.g., $l \equiv \log L$, $m_{BH} \equiv \log
M_{BH}$.

\section{The Statistical Model and Posterior Distribution}

\label{s-statmodel}

Early estimates of the BHMF were obtained by directly binning up the virial
mass estimates and applying the $1 / V_{max}$ correction, where the $1 / V_{max}$ correction
is the same as that used to estimate the luminosity function. However, BHMFs
obtained in this manner suffer from both incompleteness caused by the
sample flux limit and artificial broadening caused by the statistical error in
the virial mass estimates \citep{kelly_bechtold2007,shen08,kelly2009}. The
incompleteness arises because there is a large range in luminosity at fixed
$M_{BH}$, scattering some quasars at fixed $M_{BH}$ above the flux limit and
some below. The artificial broadening arises because the statistical error in
the virial mass estimates scatters more quasars into bins of higher $M_{BH}$
than lower $M_{BH}$ when the BHMF declines toward higher values of $M_{BH}$.
In order to account for these effects, \citet{kelly2009} developed a Bayesian
technique for estimating both the BHMF and BHERF of Type 1 quasars which
corrects for incompleteness and the effects of statistical error in the
virial estimates; they used their method to estimate the local BHMF of Type 1
quasars. \citet{schulze2010} developed and used a similar method to estimate
the local BHMF and BHERF of Type 1 quasars, although they did not correct for
the error in the virial mass estimates. Subsequent work presented in
K10 and Paper 1 has improved upon the model of \citet{kelly2009}.

In this work we expand on the statistical model described in Paper
1. \citet{kelly2009} modeled the distribution of Type 1 quasars in the
mass-redshift plane as a mixture of log-normal distributions, and the
Eddington ratio distribution at fixed $M_{BH}$ as a single log-normal
distribution whose geometric mean depended linearly on $\log
M_{BH}$. They assumed that the mass estimates were
unbiased. \citet{kelly2010} expanded this model and used a mixture of
log-normals for the Eddington ratio distribution at fixed $M_{BH}$. In
both \citet{kelly2009} and \citet{kelly2010} the Eddington ratio
distribution was assumed to not evolve, while in Paper 1 we
incorporated evolution by estimating the BHMF and BHERF independently
in different redshift bins. In this section we describe our expansion
to the model of Paper I and summarize the important aspects of our
statistical model; further details can be found in Paper I and
\citet{kelly2009}.

\subsection{Mixture of Log-Normal Functions Model}

\label{s-lognormals}

Motivated by the observed small statistical scatter in the mass estimates for
the SDSS, Paper I expanded on the model of \citet{kelly2009} to incorporate a
more flexible model for the error distribution of the mass
estimates. In both Paper I and this paper the error distribution is
modeled as a Gaussian distribution with unknown variance and
optionally and unknown mean. Because 
we use FWHM-based virial mass estimates, our model is only with respect to
these mass estimates; mass estimates based on the line dispersion or other
line width measures may have a different error distribution. Paper I assumed
that the mass estimates were unbiased at the average luminosity as a function
of $M_{BH}$, but that the mass estimates potentially exhibited a
luminosity-dependent bias at fixed mass. They assumed the following model for
the mass estimates:
\begin{equation}
  m_{vir} = m_{BH} + \beta [l - E(l|m_{BH})] + \sigma_{ml} \epsilon_{ml}.
  \label{eq-mvirial}
\end{equation}
Here, $m_{vir} \equiv \log M_{vir}, m_{BH} \equiv \log M_{BH}, l
\equiv \log L$, $E(l|m_{BH})$ is the expectation value of $\log L$ at fixed
$M_{BH}$ and $\epsilon_{ml}$ is a random variable drawn from the
standard normal distribution. The term $\beta$ models how the systematic
error (i.e., the luminosity-dependent bias) in the mass estimates
scales with luminosity, and $\sigma_{ml}$ is the standard deviation in
the mass estimates at fixed $M_{BH}$ and $L$. The standard mass
estimates assume $\beta = 0$ and $\sigma_{ml} \sim 0.4$ dex.

\begin{figure*}
  \begin{center}
  \includegraphics[scale=0.3,angle=90]{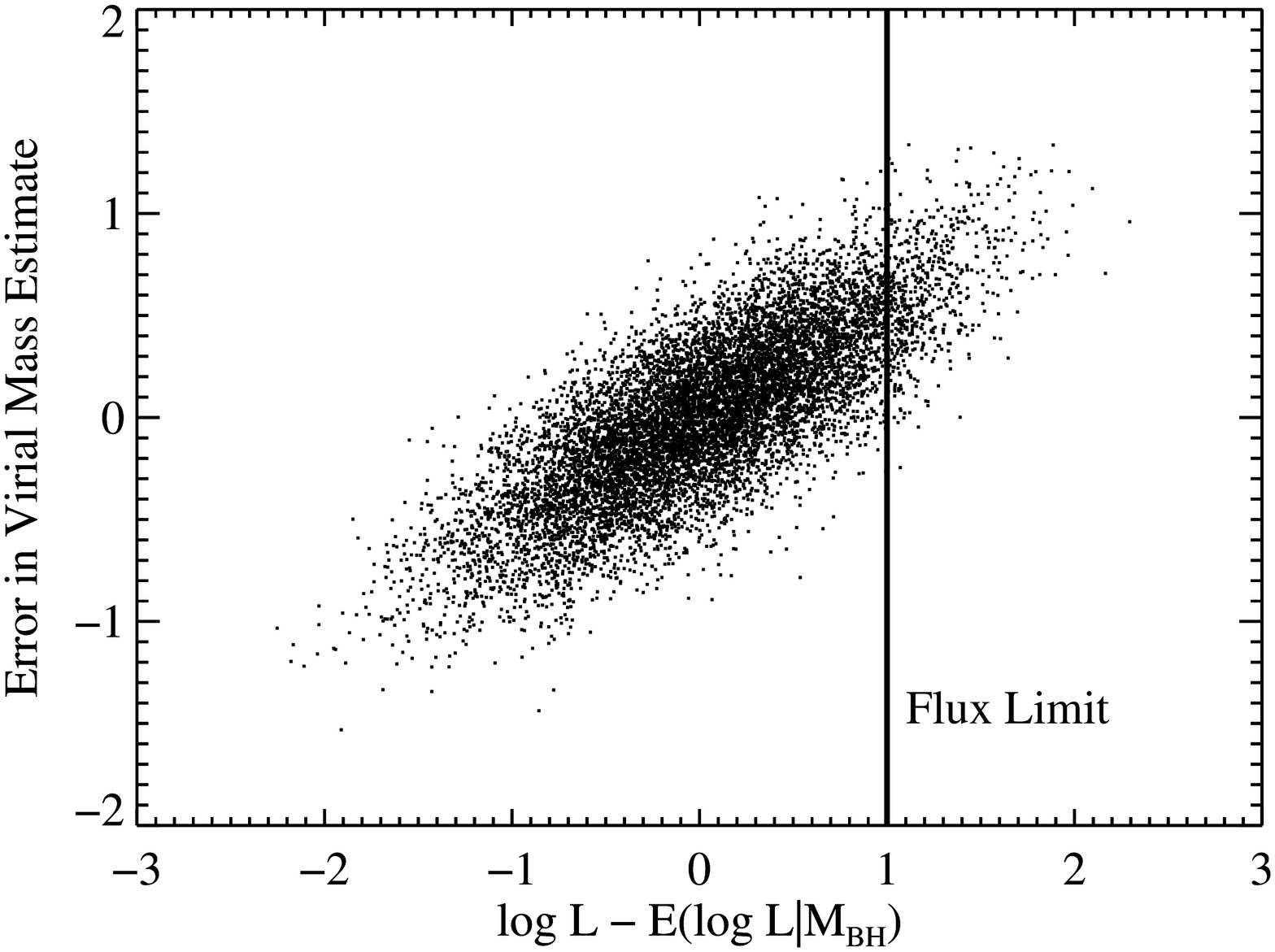}
  \includegraphics[scale=0.3,angle=90]{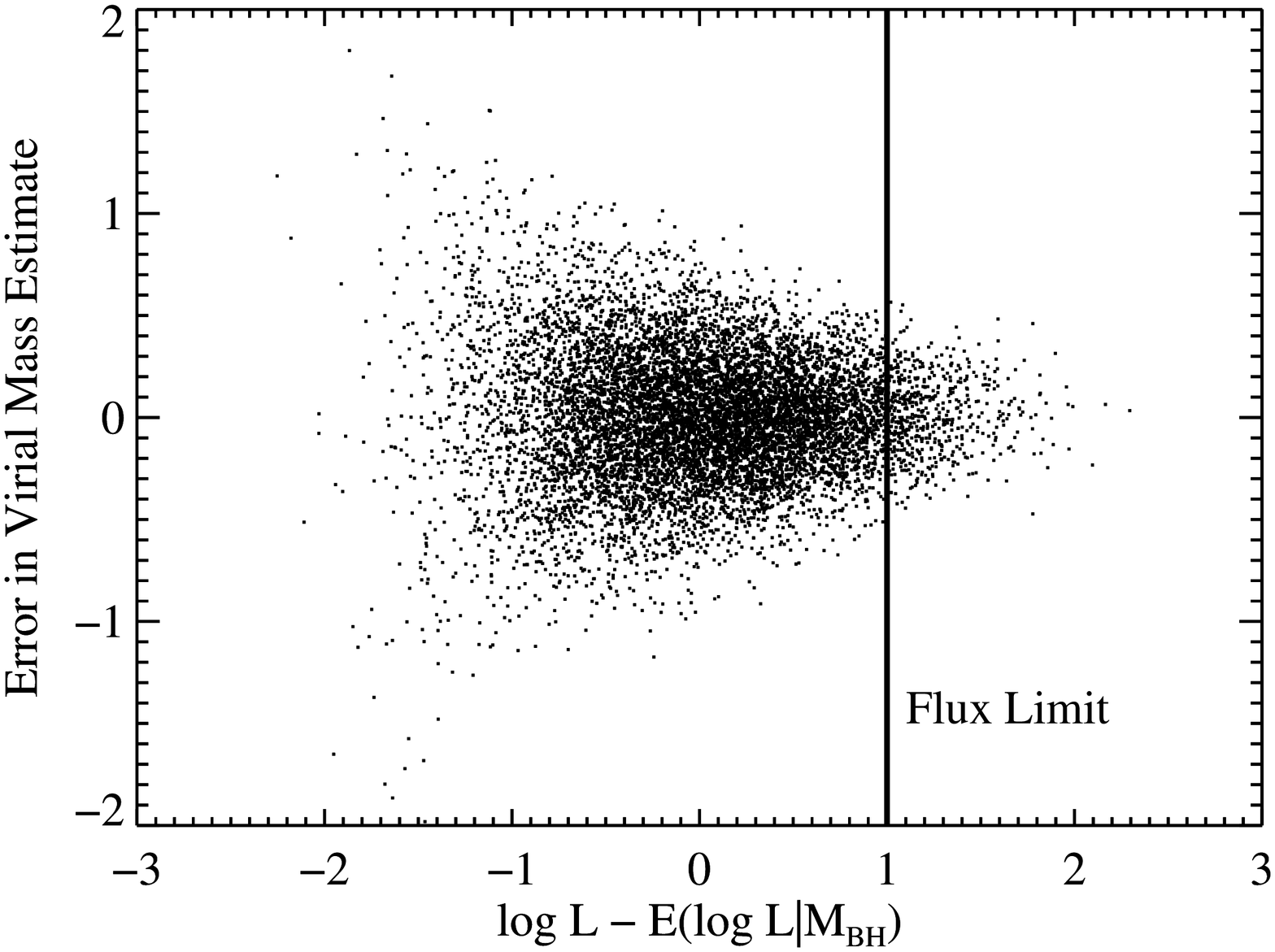}
  \caption{Illustration of the two models for the distribution of
    virial mass estimate errors used in this work. Mock data are generated
    using Eqn.\ (\ref{eq-mvirial}). In both panels $E(\log L|M_{BH})$
    denotes the mean value of $\log L$ at fixed $M_{BH}$. The left panel
    shows a model with a luminosity dependent bias ($\beta > 0$)
    and constant $\sigma_{ml}$ with luminosity, while the right
    panel shows a model with $\beta=0$ and a decrease in the
    amplitude of the error in the mass estimates toward higher $L$. In
    both models the dispersion in virial estimates decreases when limited
    to only those quasars in the bright tail of the luminosity
    distribution. However, the virial mass estimates are also biased if
    the small dispersion in mass estimates observed in the SDSS is caused
    by the situation depicted in the left panel ($\beta > 0$).}
  \label{f-merr_illust}
\end{center}
\end{figure*}

The motivation for modeling the mass estimates according to Equation
(\ref{eq-mvirial}) is illustrated in Figure \ref{f-merr_illust}. Here we show
the distribution of the error in the mass estimates at fixed true mass as a
function of luminosity, assuming a value of $\beta = 0.5$ and $\sigma_{ml} =
0.25$. When averaging over a broad range of luminosity, as is done for the
reverberation mapping sources, the distribution of the mass estimate errors
is broad. However, when limiting ourselves to a narrow luminosity range at
the bright end, as in a flux-limited sample like the SDSS, the mass estimates
exhibit a bias and smaller scatter. This model is thus one way of reconciling
the fact that the scatter in the mass estimates for the SDSS imply smaller
uncertainties in $M_{vir}$ than does the scatter in the mass estimates for
the reverberation mapping sample. Some physical reasons to expect
such a luminosity-dependent bias were further discussed in \S3.2.1 of Paper
I. Alternatively, another possibility is that the mass estimates at fixed
mass and luminosity are always unbiased (i.e., $\beta = 0$), but
$\sigma_{ml}$ decreases toward higher luminosity. This possibility is also
illustrated in Figure \ref{f-merr_illust}. Currently it is not possible to
unambiguously distinguish between these two possibilities solely
from AGN demographics, and both situations may be at work. As discussed in
Paper I, there is some indication from the reverberation mapping data for NGC
5548 that $\beta > 0$ for FWHM-based virial mass estimates, although this
result is only moderately significant at $2.4\sigma$. As such, in this work
we obtain constraints on the BHMF and Eddington ratio distribution under each
of these two models for the error in the mass estimates.

The values of $\beta$ and $\sigma_{ml}$ are assumed to be different
when different emission lines are used to calculate $M_{vir}$. In this work
we use the \hbeta, \MgII, and \CIV\ emission lines. These lines have
different advantages and disadvantages, and the behavior of the mass
estimates derived from the \MgII\ and \CIV\ lines are less well understood
than those derived from \hbeta. We refer the reader to Paper I for a
discussion of the issues and possible biases surrounding virial mass
estimates derived from different emission lines \citep[see also][and the
discussion in \S~\ref{s-systematics}]{shen_liu2012}.

In this work we model the joint distribution of $M_{BH}$ and $L$ for Type 1
quasars in a given redshift bin as a mixture of $K$ 2-dimensional
log-normal distributions:
\begin{align}
\lefteqn{p(m_{BH},l|\pi,\mu,\Sigma) =} \nonumber \\
& \sum_{k=1}^K \frac{\pi_k}{2\pi |\Sigma_k|^{1/2}} \exp\left\{-\frac{1}{2}({\bf x} -
\mu_k)^T \Sigma_k^{-1} ({\bf x} - \mu_k)\right\}.
  \label{eq-bhmf}
\end{align}
Here, ${\bf x} = [m_{BH}, l]^T$, ${\bf x}^T$ denotes the transpose of ${\bf
x}$, $\pi = (\pi_1,\ldots,\pi_K)$, $\mu=(\mu_1,\ldots,\mu_K)$, and $\Sigma =
(\Sigma_1,\ldots, \Sigma_K)$. The terms $\pi$ denote the relative
contribution of each 2-dimensional Gaussian function to the model
distribution, the terms $\mu$ denote the means of each Gaussian
function in the $M_{BH}$--$L$ plane, and the terms $\Sigma$ denote the
covariance matrices of each Gaussian function in the $M_{BH}$--$L$
plane. Note that Equation (\ref{eq-bhmf}) is with respect to the true
values of $M_{BH}$, and not with the respect to the virial mass
estimates; Equation (\ref{eq-mvirial}) describes how to connect the
distribution of the virial mass estimates to Equation
(\ref{eq-bhmf}). 

As in Paper I we model the distribution of $z$ 
in each redshift bin as a Pareto distribution (i.e., a power-law):
\begin{equation} p(z|\gamma) = \frac{(1 + \gamma) z^{\gamma}}{z_{\rm
max}^{1+\gamma} - z_{\rm min}^{1 + \gamma}}
  \label{eq-zdist}.
\end{equation}
In this work we use a value of $K = 5$, which provides considerable
flexibility. There was little difference in employing larger values of $K$ and such larger values did not justify the increased computational burden. Note that the comoving number density per 3-dimensional
box defined by $(m_{BH}, l, z)$ and $(m_{BH}, l,z) + (dm_{BH}, dl, dz)$ is
\begin{equation}
  \phi(m_{BH}, l, z) = N \left(\frac{dV}{dz}\right)^{-1}
  p(m_{BH},l|\pi,\mu,\Sigma) p(z|\gamma), \label{eq-ndensity}
\end{equation}
where $N$ is the total number of Type 1 quasars in the observable
universe, and $dV / dz$ is the derivative of comoving volume with
respect to redshift. The mass or luminosity function is obtained by
integrating Equation (\ref{eq-ndensity}) over $l$ or $m_{BH}$,
respectively.

In the absence of measurement error, Equations
(\ref{eq-mvirial})--(\ref{eq-zdist}) provide us with all of the necessary
ingredients to construct the likelihood function; i.e., these equations tell
us how to connect the mass and luminosity functions to the observed
distribution of virial mass estimates, luminosities, and redshifts. However,
in reality the line widths and luminosities are measured with error, and
therefore so are the virial mass estimates. In general, the measurement error
on the FWHM dominates over the measurement error on the luminosity, so we
only account for measurement errors on the virial mass estimates. Denote the
measured virial mass estimate as $\hat{M}_{vir}$ ($\hat{m}_{vir} \equiv \log
\hat{M}_{vir}$). In this work we model the measurement error distribution of
the virial mass estimates using a Student's $t$-distribution centered at the
true value of $M_{vir}$:
\begin{align}
  \lefteqn{ p(\hat{m}_{vir}|m_{vir}) =} \nonumber \\
& \frac{\Gamma( (\nu + 1) / 2
    )}{\sigma_{\rm meas} \Gamma(\nu / 2) \sqrt{\nu \pi}} \left[ 1 +
    \frac{1}{\nu} \left( \frac{\hat{m}_{vir} - m_{vir}}{\sigma_{\rm
          meas}} \right)^2 \right]^{-(\nu + 1) / 2} \label{eq-tdist}.
\end{align}
Here, $\nu$ is the degrees of freedom of the $t$-distribution, $\sigma_{\rm
meas}$ is the fixed measurement error amplitude, which is calculated from the
emission line fitting procedure, and $\Gamma(\cdot)$ is the Gamma function. We use the
student's $t$-distribution because it is considered a robust alternative to
the normal distribution. In the limit $\nu \rightarrow \infty$ the
$t$-distribution converges to the normal distribution, but for finite $\nu$
the $t$-distribution has heavier tails. As a result, the $t$-distribution
downweights the influence of outlying values of the virial mass estimates,
which may be caused by systematics involving bad line width measurements, or
due to the presence of a population of AGN for which the virial mass
estimates are subject to a large systematic error. In this work we use $\nu =
8$, which is a typical value for robust analysis; there is little
difference in using similar values of $\nu$. This is a slight
improvement over Paper I, where we modeled the 
measurement error distribution as a normal distribution.

For a given redshift bin, denote the set of parameters for our statistical
model as $\theta = (\pi, \mu, \Sigma, \gamma, \beta, \sigma_{ml})$, the
vector of measured virial masses as $\hat{\bf M}_{vir}$, the vector of
measured luminosities as ${\bf L}$, the vector of measured redshifts as
${\bf z}$, and the number of Type 1 quasars in our sample as $n$. Following
\citet{kelly2009}, we derive the posterior distribution from Equations
(\ref{eq-mvirial})--(\ref{eq-zdist}) and (\ref{eq-tdist}) as
\begin{align}
\lefteqn{p(\theta|\hat{\bf M}_{vir}, {\bf L}, {\bf z}) =
p(\theta) [p(I=1|\theta)]^{-n}} \nonumber \\
& \times \prod_{i=1}^n p(z_i|\gamma) \int_{-\infty}^{\infty}
p(\hat{m}_{vir,i}|m_{vir,i}) p(m_{vir,i},l_i|\theta) dm_{vir,i},
\label{eq-post}
\end{align}
where
\begin{align}
\lefteqn{p(m_{vir,i}, l_i|\theta)  =} \nonumber \\
& \sum_{k=1}^K \frac{\pi_k}{2\pi |V_k|^{1/2}} \exp\left\{-\frac{1}{2}({\bf x}_{vir,i} - \mu_k)^T V_k^{-1} ({\bf
x}_{vir,i} - \mu_k)\right\} \label{eq-likhood}
\end{align}
\begin{equation}
{\bf x}_{vir,i}  =  [m_{vir,i}, l_i]^T \label{eq-xvir}
\end{equation}
\begin{equation}
V_k  =  \left( \begin{array}{cc} Var(m_{vir}|k) & Cov(m_{vir},l|k) \\ Cov(m_{vir},l|k) & \Sigma_{L,k}
    \end{array} \right) \label{eq-vmat1}
\end{equation}
\begin{equation}
Var(m_{vir}|k)  =  \Sigma_{M,k} + \beta^2 (\Sigma_{L,k} - \Sigma^2_{ML,k} / \Sigma_{M,k})
+ \sigma^2_{ml} \label{eq-varmvir}
\end{equation}
\begin{equation}
Cov(m_{vir},l|k)  =  \Sigma_{ML,k} + \beta (\Sigma_{L,k} -
\Sigma^2_{ML,k} / \Sigma_{M,k}) \label{eq-covmvir}.
\end{equation}
Here, $p(\theta)$ is the prior on $\theta$, $n$ is the total number of data
points in the redshift bin, $p(I=1|\theta)$ is the probability of a Type 1
quasar from the redshift bin of interest making it into our sample, $V_k$ is
the covariance matrix of $\log M_{vir}$ and $\log L$ for the $k^{\rm th}$
log-normal function, and $\Sigma_{M,k}, \Sigma_{L,k},$ and $\Sigma_{ML,k}$
denote the variance in $\log M_{BH}$, variance in $\log L$, and covariance
between $\log M_{BH}$ and $\log L$ for the $k^{\rm th}$ log-normal function,
respectively. Equation (\ref{eq-likhood}) is obtained by averaging
Equation (\ref{eq-bhmf}) over the distribution of $m_{vir}|m_{BH},l$ implied by
Equation (\ref{eq-mvirial}) with respect
to $m_{BH}$. The parameters, $\theta$, are estimated independently in each
redshift bin.

As in Paper I, we define $L$ to be the luminosity at 2500\AA, which we derive
from the $i$-band magnitude according to the prescription given in
\citet{richards2006}. The term $p(I=1|\theta)$ is the probability of
including a Type 1 quasar in our sample given the model luminosity
function, and is calculated from the SDSS 
selection function, $s(L,z)$, as
\begin{equation} p(I=1|\theta) = \int_{0}^{\infty} \int_{0}^{\infty}
s(L,z) p(L|\pi,\mu,\Sigma) p(z|\gamma)\ dL\ dz.
\end{equation}
The selection function of our sample is the same as that for the sample
in \citet{richards2006}, given in that paper.

\subsection{The Prior Distribution}

\label{s-prior}

Flexible models such as the mixture of log-normal functions model we use
enable modeling of a broad range of distributions, but they can also suffer
from overfitting because the data do not provide enough information on the
structure of the distributions. It is often useful to impose priors invoking
smoothness on the estimated distributions, otherwise highly `wiggly' BHMFs
are considered just as likely as smooth ones \emph{a priori}. In addition, because our sample
is truncated, there is little to no information from the data on the distribution in the
$M_{BH}$--$L$ plane below the flux limit. Therefore, it is necessary to
impose several constraints on $\theta$ via the prior distribution in order to
keep the solution from wandering into unreasonable regions.

The prior constraints that we impose are as follows:
\begin{itemize}
  \item The standard deviation of each log-normal function must be
      between 0.2 and 1.0 dex for both $\log M_{BH}$ and $\log L / L_{Edd}$.
  \item The mean of $\log M_{BH}$ for each log-normal function must be
      between 6.0 and 10.0. This reflects our assumption that the BHMF must decrease at $M_{BH} < 10^6 M_{\odot}$ and $M_{BH} > 10^{10} M_{\odot}$.
  \item The mean value of $\log L / L_{Edd}$ for each log-normal function
      must be between -3.0 and 0.0. The lower bound was chosen because
      accretion flows are thought to undergo a state-transition near $L /
      L_{Edd} \sim 10^{-2}$--$10^{-3}$, and the broad line region is not
      expected to exist below this critical Eddington ratio
      \citep[e.g.,][]{czerny2004,hhqh09,trump2011}. The upper bound was chosen to reflect our assumption that the BHERF must decrease above the Eddington limit.
  \item The fraction of Type 1 quasars in each log-normal function
      radiating at $\log L / L_{Edd} < -3.5$ must be less than
      $1\%$. This constraint was added in addition to the constraint
      on the mean value of $\log L / L_{Edd}$ to ensure that the
      estimated distribution of Type 1 quasars declined strongly at
      $L / L_{Edd} \lesssim 10^{-3}$.
\end{itemize}

In order to test how sensitive our results are to these prior
constraints, we have also performed the analysis allowing the standard
deviation for both $\log M_{BH}$ and $\log L / L_{Edd}$ to be as high
as 2.0 dex, and extending the lower limit for the mean value of $\log
L / L_{Edd}$. Extending the upper bound on $\Sigma_{M,k}$ does not
significantly change our results beyond slightly increasing the
uncertainties, with the only exception being that derived value for the
maximum value of $M_{BH}$ in the observable universe for a Type 1
quasar is very sensitive to this upper bound; this is discussed
further in Section \ref{s-most_massive}. In addition, the shape of the
BHMF and BHERF at $M_{BH} \gtrsim 3 \times 10^8 M_{\odot}$ and $L / L_{Edd}
\gtrsim 0.05$ are not sensitive to the lower bound on the mean value
of $\log L / L_{Edd}$. However, the normalization of the BHMF and
BHERF are sensitive to this lower bound, as this lower bound is
directly related to how many Type 1 quasars are sample is
missing. Luckily this is not a problem for our analysis as none of our
scientific conclusions depend on the absolue normalization of the BHMF
and BHERF.

Beyond the bounds listed above, we use a prior for the log-normal parameters
that is based on the `partially-proper
prior' of \citet{roeder97}. We constrain the $K$ log-normal functions by
ordering their values of $\mu_k$ by increasing mean luminosity, i.e., the $k
= 1$ log-normal function has the faintest mean luminosity, and the $k = K$
log-normal function has the brightest. In addition, denote $\mu_{L,k}$ and
$\mu_{M,k}$ to be the components of $\mu_k$ corresponding to the mean $\log
L$ and $\log M_{BH}$, respectively. Our prior for the luminosity components
of $\theta$ is
\begin{equation}
p(\mu_{L},\Sigma_{L}) = p(\mu_{L,1}) \prod_{k=2}^K
p(\mu_{L,k},\Sigma_{L,k}|\mu_{L,k-1},\Sigma_{L,k-1}) \label{eq-lprior}
\end{equation}
\begin{equation}
p(\mu_{L,1}) = \frac{\mu_{L,1} - l_{\rm min}}{l_{\rm min} -
l_{\rm max}} \left(1 - \frac{\mu_{L,1} - l_{\rm min}}{l_{\rm min} -
l_{\rm max}}\right)^{K-1}  \label{eq-lprior1}
\end{equation}
\begin{align}
\lefteqn{p(\mu_{L,k},\Sigma_{L,k}|\mu_{L,k-1},\Sigma_{L,k-1}) =} \nonumber \\
& \frac{1}{2\pi h(k,k-1)} \left[\Phi\left(\frac{l_{\rm max} -
\mu_{L,k-1}}{h_L(k,k-1)}\right) - \frac{1}{2}\right]^{-1} \nonumber \\
& \times N(\mu_{L,k}|\mu_{L,k-1},h_L(k,k-1)) H(\mu_{L,k} - \mu_{L,k-1}) \label{eq-lpriork}
\end{align}
\begin{equation}
h_L(k,k-1) = 2 \left(\frac{1}{\Sigma_{L,k}} +
\frac{1}{\Sigma_{L,k-1}}\right)^{-1} \label{eq-lhmean}.
\end{equation}
Here, $l_{\rm min}$ and $l_{\rm max}$ are the minimum and maximum possible
values of $\mu_{L,k}$, $h_L(k,k-1)$ denotes the harmonic mean of
$\Sigma_{L,k}$ and $\Sigma_{L,k-1}$, $\Phi(\cdot)$ is the standard normal
cumulative distribution function, $N(x|\mu,V)$ denotes a Gaussian
function in $x$ with mean $\mu$ and variance $V$, and $H(\cdot)$ is
the Heaviside step function. Given our prior
constraints, $l_{\rm min}$ is the value of $\log L$ for $M_{BH} = 10^6
M_{\odot}$ and $L / L_{Edd} = 10^{-3}$, and $l_{\rm max}$ is the value of
$\log L$ for $M_{BH} = 10^{10} M_{\odot}$ and $L / L_{Edd} = 1$. We chose the
prior on $\mu_{L,1}$ because it is the probability distribution for the
minimum value of a sample of $K$ random variables uniformly distributed
between $l_{\rm min}$ and $l_{\rm max}$. The prior on $\mu_{L,k}$ for $k > 1$
was chosen to enforce smoothness in the estimated luminosity function, as it
places higher probability on solutions where the individual log-normal
functions are close together with respect to their average standard
deviations. Our prior on $\mu_{L}$ was therefore constructed to not favor a
particular value for the centroid of the luminosity function, under the
constraints listed above, but to favor smooth luminosity functions.

We also place a similar `smoothness' prior on $\mu_{M}$, but without
enforcing a particular ordering of the log-normal functions:
\begin{equation}
p(\mu_{M}|\Sigma_M,\bar{m}) = \prod_{k=1}^K
\frac{1}{\sqrt{2\pi h_{M}}}\exp\left\{-\frac{1}{2}\frac{(\mu_{M,k} -
\bar{m})^2}{h_M}\right\} \label{eq-mprior}
\end{equation}
\begin{equation}
  h_M = K \left( \sum \frac{1}{\Sigma_{M,k}}
  \right)^{-1} \label{eq-mhmean}
\end{equation}
The `hyper-parameter' $\bar{m}$ is the mean value of
the $\{\mu_{M,k}\}$ and is an additional parameter. We assume a uniform prior on
$\bar{m}$ over all possible values; note that even though we do not
place explicit bounds on $\bar{m}$, such bounds are implied by the
bounds placed on $\mu_{M,k}$. We place
uniform priors on $\Sigma_{L,k}$ and $\Sigma_{M,k}$ under the constraints
given above. We obtain the prior on $\Sigma_{ML,k}$ by placing a uniform
prior on the angle that the slope of the mean value of $\log L / L_{Edd}$ as
a function of $M_{BH}$ makes with the horizontal. For $\pi$, we assume a
Dirichlet prior with parameter $\alpha = 0.1$. Finally, we use the same prior
on $\beta$ and $\sigma_{ml}$ as that described in \S~3.2.1 of Paper I, which
is constructed to give results consistent with the reverberation mapping
sample.

\section{The Black Hole Mass Function}
\label{s-bhmf}

We applied our Bayesian method to our sample derived from the SDSS DR7 data
set for both the model with a luminosity-dependent bias ($\beta \neq 0$) and
the model with decreasing scatter in the mass estimates at higher luminosity
($\beta = 0$). For the latter model  ($\beta = 0$) we do not
explicitly fit for a dependence of $\sigma_{ml}$ on $L$, but rather
just assume a single value of $\sigma_{ml}$ over the luminosity range
in each redshift bin. This should be an adequate approximation as the
luminosity range of our data set in each redshift bin is
narrow. Details of the sample are described in Paper I

Assuming the model with a luminosity-dependent
bias, we estimate the slopes of the bias to be $\beta_{\rm H\beta} \approx
0.1 \pm 0.1, \beta_{\rm MgII} \approx 0.39 \pm 0.13,$ and $\beta_{\rm CIV}
\approx 0.40 \pm 0.11$ for the \hbeta,\MgII, and \CIV\ emission lines,
respectively. For the $\beta \neq 0$ model we estimate the scatter in the mass estimates at fixed
luminosity and BH mass to be $\sigma_{\rm H\beta} \approx 0.24
\pm 0.03, \sigma_{\rm MgII} \approx 0.25 \pm 0.02$, and $\sigma_{\rm CIV}
\approx 0.21 \pm 0.01$ for the \hbeta,\MgII, and \CIV\ emission lines,
respectively. These values are significantly different from the traditionally
assumed values of $\beta = 0$ and $\sigma_{BL} = 0.4$ dex. However, we note
that at $z \gtrsim 3.5$ the derived values for the \CIV\ line become
$\beta_{\rm CIV} \approx 0.3 \pm 0.3$ and $\sigma_{\rm CIV} \approx 0.39 \pm
0.04$, probably reflecting the broader range in luminosity at these redshifts
due to the deeper SDSS flux limit. This increase in $\sigma_{ml}$ as the range in luminosity is
increased supports our hypothesis that the dispersion in the virial mass
estimate error is luminosity-dependent.

These results on the luminosity-dependent bias model are similar as that
obtained in Paper I using a simpler model for the Eddington ratio
distribution, showing that many of the results of Paper I are robust against
the assumed Eddington ratio distribution. We therefore conclude from this
that either the statistical scatter in the FWHM-based broad line mass estimates is
correlated with luminosity, producing a luminosity-dependent bias, or that
the scatter in the mass estimates is smaller for the luminous quasars probed
by the SDSS, or some combination of these two possibilities. In Paper I we
have already discussed the implications of these results for the broad line
mass estimates, and, as the focus of this paper is on the Type 1 quasar
black hole mass function and Eddington ratio distribution, we merely note
here that our earlier conclusions on the statistical properties of the broad
line mass estimate errors are unchanged using a more flexible model for the
Eddington ratio distribution.

For the sake of brevity, in the remainder of this work we present the results
obtained from the model that assumes that there is no luminosity-dependent
bias, i.e., $\beta = 0$. We do this for easier comparison with other work, as
almost all studies implicitly assume that there are no systematic
trends in the mass estimate errors with luminosity. However, we also discuss
what aspects of our results change when we allow the mean value of the error in the mass
estimates to depend on luminosity.

\subsection{Evaluating the Fit Quality}
\label{s-fit_quality}

In order to evaluate whether our model provides a good fit to the data we use
a technique called posterior predictive check
\citep{rubin81,rubin84,gelman96}. For each random draw of the BHMF, the
BHERF, and the values of $\sigma_{ml}$ from their joint posterior probability
distribution we generate a mock sample of virial mass estimates and
luminosities, conditional on the SDSS quasar selection function. These mock
samples are then compared to the actual data to check for consistency. This
approach therefore includes our uncertainty in the BHMF and BHERF, as well as
the randomness in generating a single sample from the BHMF and BHERF.

\begin{figure}
  \begin{center}
    \scalebox{0.9}{\rotatebox{90}{\plotone{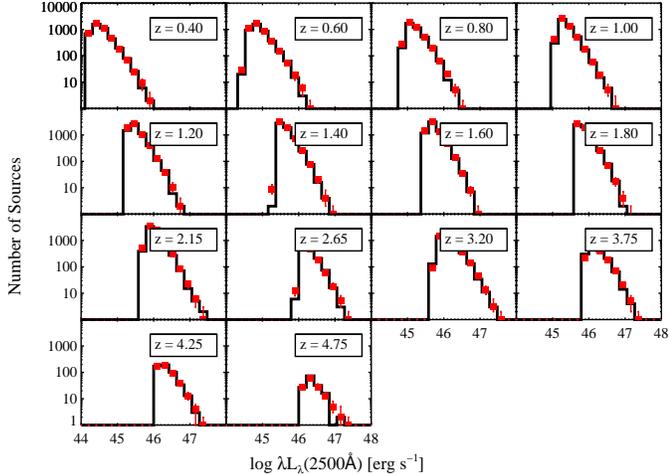}}}
    \caption{Posterior predictive check comparing the actual
      distribution of luminosities for our sample (solid histogram)
      with the set of distributions generated by our black hole mass
      and Eddington ratio functions (red squares with error
      bars). The error bars contain $68\%$ of the posterior
      probability. The distributions of luminosities generated by our
      model are consistent with the observed distributions, showing
      that our model provides an acceptable fit. \label{f-lum_check}}
  \end{center}
\end{figure}

\begin{figure}
  \begin{center}
    \scalebox{0.9}{\rotatebox{90}{\plotone{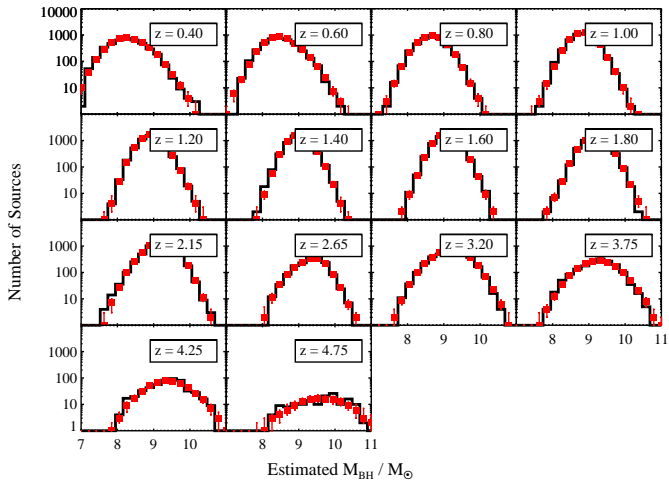}}}
    \caption{Same as Figure \ref{f-lum_check}, but for the virial mass
      estimates from \citet{shen2011}. The distributions of virial masses generated by our
      model are consistent with the observed distributions, showing
      that our model provides an acceptable fit. \label{f-mass_check}}
  \end{center}
\end{figure}

In Figures \ref{f-lum_check} and \ref{f-mass_check} we compare the histograms
of the luminosities and virial mass estimates for our sample with the
histograms of the mock samples generated from our MCMC output. Our estimated
BHMF and BHERF generate samples of the virial mass estimates and luminosities
that are consistent with our actual data, showing that our model provides an
acceptable fit. The mock samples generated by the model that includes a
luminosity-dependent bias term were also consistent with the data.

\subsection{Comoving Number Densities as a function of $M_{BH}$ and $z$}
\label{s-mbh_ndens}

\begin{figure*}
  \begin{center}
    \scalebox{0.7}{\rotatebox{90}{\plotone{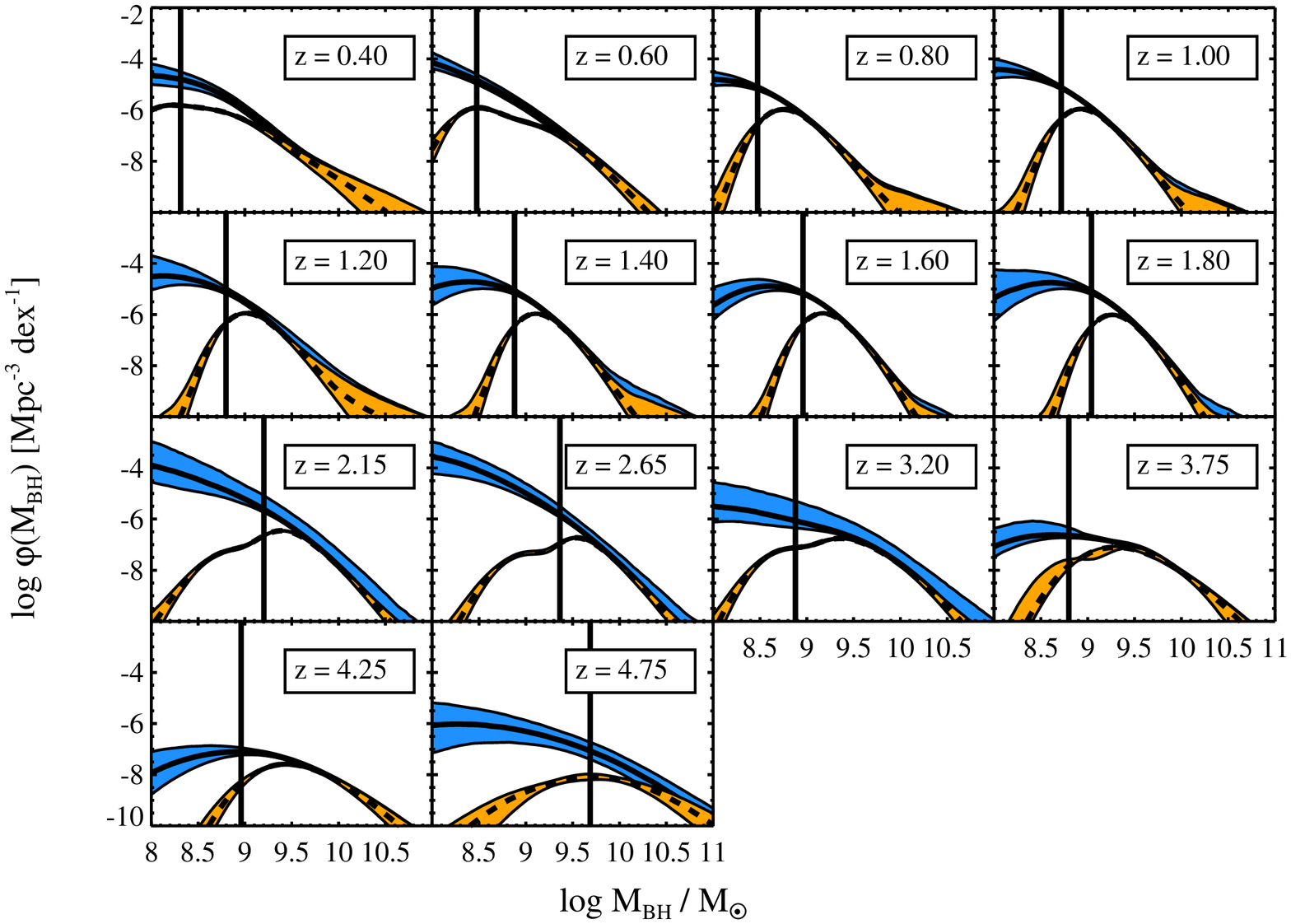}}}
    \caption{The black hole mass function for broad line AGN derived from our
      sample. The blue region contains $68\%$ of the probability for
      the BHMF, and the solid line running through it denotes the
      best-fit BHMF, defined to be the posterior median. The orange
      shaded region containes $68\%$ of the probability on the BHMF
      for Type 1 quasars above the SDSS flux limits, and the dashed line
      denotes the posterior median. The vertical solid line denotes
      the mass below which the completeness in the BHMF drops to
      $\lesssim 10\%$ for the flux-limited SDSS sample, and thus mass bins below this marker are
      highly incomplete and the extrapolation becomes highly
      uncertain. We do not find any evidence for a turnover in the
      Type 1 quasar BHMF above the
      $10\%$ completeness limit in $M_{BH}$. \label{f-bhmf}}
  \end{center}
\end{figure*}

Our estimated BHMF and its uncertainty in various redshift bins is shown in
Figure \ref{f-bhmf} and reported in Table \ref{t-bhmf}. The best-fit BHMF is
defined as the posterior median value of the number density as a function of
$M_{BH}$, and the uncertainty in the BHMF is defined to be the region
containing $68\%$ of the posterior probability for the number density as a
function of $M_{BH}$. Note that this represents the formal statistical
uncertainty on the BHMF; in reality, the actual uncertainty is larger due to
unaccounted for systematic errors. Also shown in Figure \ref{f-bhmf} is the
estimated $10\%$ completeness limit and the flux-limited BHMF for quasars at
$i < 19.1$ at $z < 2.9$ and $i < 20.2$ at $z > 2.9$, corresponding to the
approximate flux limits of the SDSS DR7 quasar sample.

\begin{deluxetable}{lcccc}
\tabletypesize{\scriptsize}
\tablewidth{0pt}
\tablecaption{Type 1 Quasar Black Hole Mass Functions \label{t-bhmf}}
\tablehead{$\bar{z}$
& $\log M_{BH}$
& $\log \Phi_{-}$\tablenotemark{a}
& $\log \Phi_0$\tablenotemark{b}
& $\log \Phi_{+}$\tablenotemark{c} \\
& $(M_{\odot})$
& \multicolumn{3}{c}{(Mpc$^{-3}{\rm dex}^{-1}$)}
}
\startdata
0.400 & 8.000 & -5.00 & -4.64 & -4.21  \\
0.400 & 8.100 & -5.03 & -4.67 & -4.30  \\
0.400 & 8.200 & -5.06 & -4.72 & -4.39  \\
0.400 & 8.300 & -5.09 & -4.78 & -4.50  \\
0.400 & 8.400 & -5.14 & -4.87 & -4.62  \\
0.400 & 8.500 & -5.22 & -4.97 & -4.76  \\
0.400 & 8.600 & -5.32 & -5.12 & -4.92  \\
0.400 & 8.700 & -5.47 & -5.29 & -5.09  \\
0.400 & 8.800 & -5.65 & -5.49 & -5.31  \\
0.400 & 8.900 & -5.87 & -5.72 & -5.54  \\
0.400 & 9.000 & -6.12 & -5.97 & -5.80  \\
0.400 & 9.100 & -6.40 & -6.25 & -6.08  \\
0.400 & 9.200 & -6.70 & -6.54 & -6.37  \\
0.400 & 9.300 & -7.02 & -6.84 & -6.67  \\
0.400 & 9.400 & -7.32 & -7.14 & -6.96  \\
0.400 & 9.500 & -7.61 & -7.43 & -7.25  \\
0.400 & 9.600 & -7.89 & -7.70 & -7.52  \\
0.400 & 9.700 & -8.17 & -7.96 & -7.76  \\
0.400 & 9.800 & -8.48 & -8.19 & -7.97  \\
0.400 & 9.900 & -8.79 & -8.41 & -8.16  \\
0.400 & 10.00 & -9.10 & -8.63 & -8.33  \\
0.400 & 10.10 & -9.44 & -8.85 & -8.51  \\
0.400 & 10.20 & -9.79 & -9.09 & -8.66  \\
0.400 & 10.30 & -10.1 & -9.33 & -8.82  \\
0.400 & 10.40 & -10.4 & -9.57 & -8.97  \\
0.400 & 10.50 & -10.8 & -9.83 & -9.12  \\
0.400 & 10.60 & -11.2 & -10.1 & -9.28  \\
0.400 & 10.70 & -11.5 & -10.3 & -9.43  \\
0.400 & 10.80 & -11.9 & -10.6 & -9.58  \\
0.400 & 10.90 & -12.3 & -10.9 & -9.76  \\
0.400 & 11.00 & -12.7 & -11.2 & -9.94 
\enddata

\tablenotetext{a}{Lower boundary on the region containing $68\%$ of the
  posterior probability for the BHMF, i.e., the $16^{\rm th}$
  percentile of the posterior distribution.}
\tablenotetext{b}{Posterior median for the BHMF.}
\tablenotetext{c}{Upper boundary on the region containing $68\%$ of the
  posterior probability for the BHMF, i.e., the $84^{\rm th}$
  percentile of the posterior distribution.}

\tablecomments{The full table is available in the electronic version of the
paper. Tabulated here are the results for the $\beta=0$ error model.
Results for the $\beta\neq 0$ error model are similar to those presented in
Paper I.}
\end{deluxetable}

It is apparent that our sample starts to become significantly incomplete at
$M_{BH} \lesssim 3 \times 10^8 M_{\odot}$, and any conclusions from the BHMF
at these masses become reliant on our model for the BHMF and our assumptions
for deriving it from the mass estimates. In addition, extrapolation in our
BHMF is unstable against small unidentified errors in both the SDSS selection
function and the mass estimates beyond the region where we are reasonably
complete. As discussed in K10, to zeroth-order the BHMF can be estimated as
$\phi_M(M_{BH}) \sim n(M_{BH}) / s(M_{BH})$, where $n(M_{BH})$ is the
measured number density of Type 1 quasars with mass $M_{BH}$, and $s(M_{BH})$
is the Type 1 quasar completeness as a function of $M_{BH}$. The term
$s(M_{BH})$ is calculated by averaging the SDSS selection function over the
luminosity distribution at fixed $M_{BH}$ (see Eq.[11] in \citet{kelly2009}).
The estimated BHMF is highly sensitive to both the measured number of sources
in a mass bin and the value of $s(M_{BH})$ when $s(M_{BH})$ is very small,
and thus is unstable against even small errors in $n(M_{BH})$ and the
selection function.

The Type 1 quasars in the most incomplete bins tend to be those with the
lowest $S/N$ spectra. Because the FWHM measurement can be biased in low $S/N$
spectra \citep{denney09}, the measured number densities of the most
incomplete mass bins may be incorrect due to biases in the mass estimates
derived from FWHM measurements, which would scatter objects into incorrect
mass bins. In addition, the SDSS quasar selection function is derived via
simulation by \citet{richards2006}. Aspects of this simulation which are not
representative of the quasar population (e.g., lack of a host galaxy
component) can introduce errors into the quasar selection function, which are
manifested as errors in $s(M_{BH})$. Therefore, in order to limit the impact
of systematics, we focus our discussion on the regions of parameter space
where the SDSS DR7 is $\gtrsim 10\%$ complete.

In general, the BHMF falls off approximately as a power-law toward higher
mass $M_{BH}$, with the BHMF being steeper at higher mass. The comoving
number density of SMBHs in Type 1 quasars continues to increase all the way
down to the masses corresponding to the $\approx 10\%$ completeness limits,
and thus we do not observe a peak in the BHMF from the SDSS. Instead, we
constrain the peak in the Type 1 quasar BHMF to occur at $M_{BH} \lesssim 2
\times 10^8 M_{\odot}$.

\begin{figure}
  \begin{center}
    \scalebox{0.9}{\rotatebox{90}{\plotone{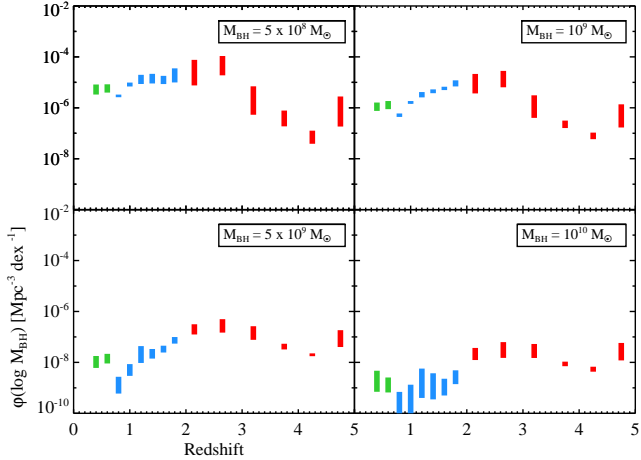}}}
    \caption{Evolution in the comoving number densities of Type 1 quasars at
      four different values of $M_{BH}$. The bars contain $68\%$ of
      the posterior probability. The green bars denote bins with
      H$\beta$-based virial mass estimates, the blue bars denote bins
      with \MgII-based mass estimates, and the red bars denote bins with
      \CIV-based mass estimates. All of the bins show a peak in the
      number density around $z \sim 2$, with the higher mass bins falling
      off more steeply toward lower redshift and less steeply toward
      higher redshift, relative to the peak. The trend is commonly
      referred to as `downsizing'. \label{f-bhmf_evol}}
  \end{center}
\end{figure}

In Figure \ref{f-bhmf_evol} we show the evolution in the Type 1 quasar BHMF
at four different values of $M_{BH}$. The number densities of SMBHs in Type 1
quasars with $M_{BH} \lesssim 10^9 M_{\odot}$ fall off more steeply toward
higher redshift at $z \gtrsim 2$, while the number densities of $M_{BH}
\gtrsim 10^9 M_{\odot}$ fall off more steeply with decreasing redshift at $z
\lesssim 2$. These trends imply that more massive black holes are more likely
to be observed at higher $z$ in Type 1 quasars than less massive ones. These
trends have been called `downsizing', and have been observed in previous work
\citep[e.g.,][K10, Paper I]{vest09}.  In addition, there is evidence for a
discontinuity in the number densities across the the redshift where we switch
from \hbeta\ to \MgII\ when calculating the mass estimates. This
discontinuity may be partly driven by systematic differences in the \hbeta-
and \MgII-based mass estimates. However, it is also likely driven at least in
part by the contribution from a host galaxy component to the nuclear emission
which boosts the number of Type 1 quasars above the flux limit at $z \lesssim
0.8$, but is not modeled in the selection function \citep{richards2006}.
Considering this, it is likely that the number densities at $z \lesssim 0.8$
are overestimated, especially at $z = 0.4$.

\begin{figure*}
  \begin{center}
    \scalebox{0.7}{\rotatebox{90}{\plotone{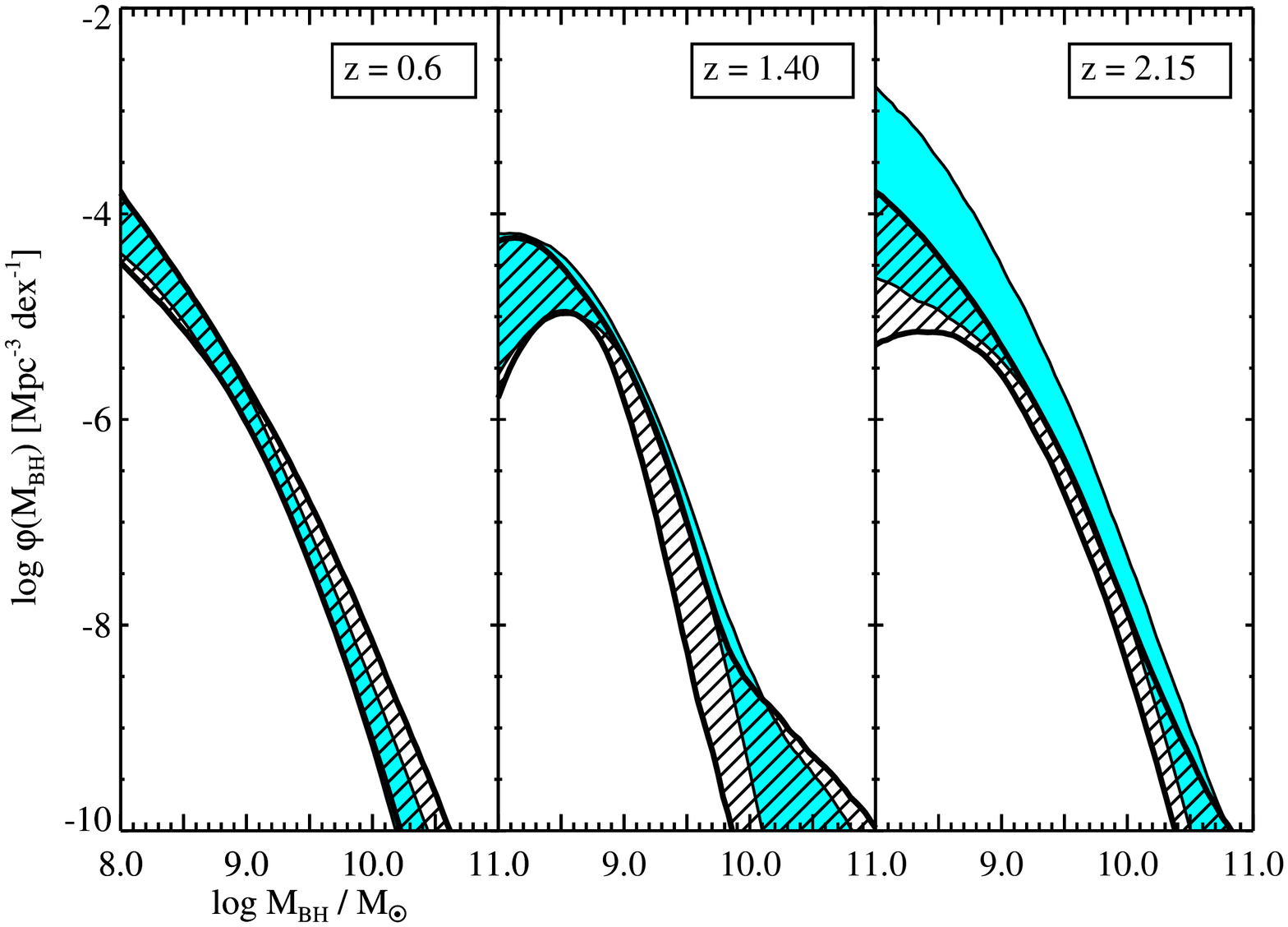}}}
    \caption{Comparison of the BHMF derived from the model with a
      luminosity dependent bias ($\beta \neq 0$, diagonal line-filled
      region) and without a luminosity dependent bias ($\beta = 0$, cyan
      region). For simplicity, we only show one redshift for each emission
      line used to calculate the virial mass estimates; the omitted bins
      exhibit similar differences. There is no apparent difference in
      the BHMFs derived from
      the \hbeta\ line, while the BHMFs derived from the \MgII\ and
      \CIV\ lines under the model with $\beta \neq 0$ are shifted lower
      in $M_{BH}$ by $\approx 0.2$ dex. This is a rather small
      difference and in general our scientific conclusions do not change
      when we allow a luminosity-dependent bias.}
    \label{f-bhmf_compare}
  \end{center}
\end{figure*}

In Figure \ref{f-bhmf_compare} we compare the mass functions derived from
both models for the error distribution in the virial mass estimates at three
representative redshifts. There is little difference in the BHMFs derived from our models
with and without a luminosity dependent bias in the mass estimates, with the
exception that the BHMF derived from \MgII\ and \CIV\ is shifted toward lower
values of $M_{BH}$ by $\approx 0.2$ dex for the model which includes a
luminosity dependent bias (i.e., $\beta \neq 0$). In addition the uncertainties in
the BHMF are larger when we allow for a luminosity-dependent bias. We
also note that the BHMF for the $\beta \neq 0$ error model is
consistent with that presented in Paper I, but with larger
uncertainties on account of the more flexible Eddington ratio distribution model.

While the derived BHMF does not depend strongly on the existence of a
luminosity-dependent bias, the flux-limited BHMF does depend strongly on the value of $\beta$.
This is because the segment of the SMBH population that is probed within a
certain luminosity range depends strongly on how the errors in the mass
estimates scale with luminosity. However, this is not a concern as the
flux-limited BHMF is not of interest in this work. However, it would be a
concern, for instance, when using virial black hole mass estimates
from flux-limited samples to investigate
models for accretion flows or evolution of the BH-bulge scaling
relation at high-redshift \citep[e.g.,][]{shen_kelly2010}.

\subsection{How Massive can Black Holes Become?'}
\label{s-most_massive}

Following K10, we compute the probability distribution of $M_{BH}^{\rm
  max}(z)$, the most massive SMBH from a population of Type 1 quasars drawn
from the BHMF within each redshift bin. In other words, this quantity may be
thought of as the most massive SMBH that would have been observed within each
redshift bin in an all-sky survey of Type 1 quasars without a flux
limit. The probability distribution for the maximum value of $M_{BH}$
generated by a random draw from the BHMF is calculated as
\begin{align}
\lefteqn{p(\log M_{BH}^{\rm max}|\pi,\mu,\Sigma,N) = N p(\log M^{\rm
    max}_{BH}|\pi,\mu,\Sigma)} \nonumber \\
& \times \left[ \int_{-\infty}^{\log M_{BH}^{\rm max}} p(m_{BH}|\pi,\mu,\Sigma)
  \ dm_{BH} \right]^{N-1}, \label{eq-maxmass}
\end{align}
where $p(m_{BH}|\pi,\mu,\Sigma)$ is obtained by marginalizing Equation
(\ref{eq-bhmf}) over luminosity. In order to incorporate our
uncertainty on the derived BHMF, we calculate Equation
(\ref{eq-maxmass}) for each value of $(N,\pi,\mu,\Sigma)$ returned by
our MCMC sampler and average the results.

\begin{figure}
  \begin{center}
    \scalebox{0.9}{\rotatebox{90}{\plotone{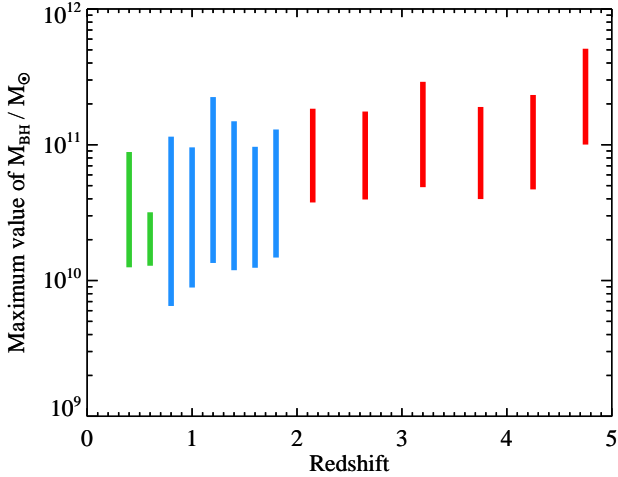}}}
    \caption{Constraints on the maximum mass that could be observed
      in a broad-line quasar as a function of redshift. Symbols are
      as in Figure \ref{f-bhmf_evol}. The maximum value of $M_{BH}$
      for quasars implied by our BHMF is $M^{\rm max}_{BH} \sim 4
      \times 10^{10} M_{\odot}$, although values of $10^{10} <
      M_{BH}^{\rm max} / M_{\odot} < 10^{11}$ are consistent within
      the uncertainties. However, as discussed in the text the values
      of $M_{BH}^{\rm max}$ we derive are highly sensitive to our
      adopted prior constraints, and the values shown should be
      treated as lower limits. \label{f-most_massive}}
  \end{center}
\end{figure}

Our derived constraints on $M_{BH}^{\rm max}(z)$ are shown in Figure
\ref{f-most_massive}. There are no obvious trends between $M_{BH}^{\rm
max}(z)$ and redshift, although there may be a slight trend for
$M_{BH}^{\rm max}(z)$ to be larger at higher $z$. This is consistent
with `downsizing', but may also be due to larger systematic errors in
\CIV-based mass estimates. The results for the luminosity-dependent
bias model are very similar, but the uncertainties are $\sim 30\%$
larger.

As mentioned in \S~\ref{s-prior} our derived constrained on
$M_{BH}^{\rm max}$ are sensitive to the prior placed on
$\Sigma_{M,k}$. This is because large value of $\Sigma_{M,k}$ imply
tails in the BHMF that extend to larger value of $M_{BH}$, and
therefore larger values of $M_{BH}^{\rm max}$. If we constrain
$\Sigma_{M,k} < 2$ instead of $\Sigma_{M,k} < 1$, then the upper
boundary of the error bars on $M_{BH}^{\rm max}$ extend to $M_{BH}
\sim 10^{12}$--$10^{13} M_{\odot}$. Ideally the prior should not have
a strong influence on this quantity, but unfortunately the upper flux
limit for the SDSS results in the extreme high mass tail of the BHMF
being poorly constrained; because of the upper flux limit, there is
nothing in the data from prohibiting a small population of Type 1
quasars hosting SMBHs with, say, $M_{BH} \sim 10^{12}
M_{\odot}$. Considering this, we can formally only place a lower bound
of $M_{BH}^{\rm max} \sim 10^{10} M_{\odot}$ on the maximum mass in
the observable universe of a SMBH in a Type 1. This is consistent with
the range $2 \times 10^{10} M_{\odot} \lesssim M_{BH}^{\rm max}
\lesssim 5 \times 10^{10} M_{\odot}$ at $z > 1$ that K10 estimated
from their sample based on the SDSS DR3.

\subsection{Completeness in Black Hole Mass}
\label{s-mass_completeness}

\begin{figure}
  \begin{center}
    \scalebox{0.9}{\rotatebox{90}{\plotone{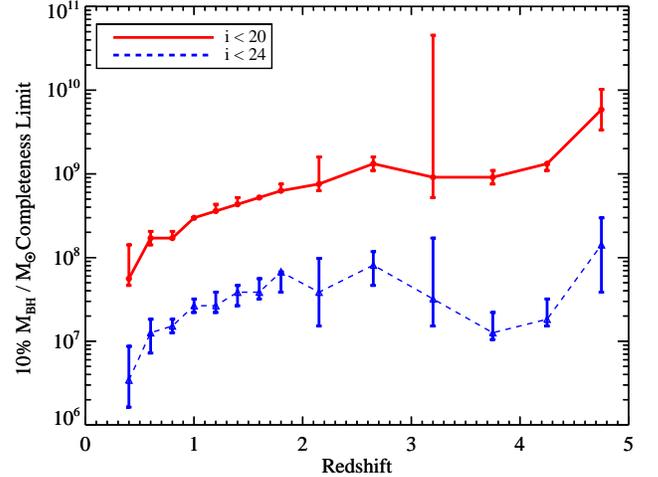}}}
    \caption{The values of $M_{BH}$ at which a Type 1 quasar survey becomes only $10\%$
      complete under a limiting
      magnitude of $i < 20$ (red solid line) and $i < 24$ (blue dashed line), as a
      function of redshift. The
      data points denote the posterior median values, and the error
      bars contain $68\%$ of the posterior probability. The SDSS
      becomes highly incomplete below $M_{BH} \sim 3 \times 10^8
      M_{\odot}$. \label{f-mass_completeness}}
  \end{center}
\end{figure}

In Figure \ref{f-mass_completeness} we show the $10\%$ completeness
limit in $M_{BH}$ as a function of $z$ for a Type 1 quasar survey with
a flux limit of $i < 20$ and $i < 24$. In calculating these
completeness limits we assume a step function down to the flux limit,
i.e., we do not assume the SDSS selection function. The $i < 20$ flux
limit roughly corresponds to the SDSS flux limit, while the $i < 24$
flux limit is similar to the limiting magnitude for the
\emph{Pan-STARRS} Medium Deep Survey Fields \citep{saglia2012} as well
as the spectroscopic samples from \emph{COSMOS}
\citep[e.g.,][]{lilly2007,trump07}. As with other quantities derived in this
section, the results from the model with a luminosity-dependent bias
in the mass estimates are similar but with larger
uncertainties. Samples with a limiting magnitude of $i = 20$ start to
become strongly incomplete at $M_{BH} \sim 10^8 M_{\odot}$ by $z \sim
1$, increasing to $M_{BH} \sim 7 \times 10^8 M_{\odot}$ by $z \sim
2$. However, Type 1 quasar samples with a limiting magnitude of $i =
24$ are able to go an order of magnitude `deeper' in $M_{BH}$,
becoming incomplete at $M_{BH} \sim 10^7 M_{\odot}$ by $z \sim 1$ and
$M_{BH} \sim 7 \times 10^7 M_{\odot}$ by $z \sim 2$.

The contribution from the host-galaxy is neglected in these calculations, but
would likely become important near $i \sim 24$, at least at lower redshift.
The host-galaxy contribution would affect these calculations in two ways.
First, the host galaxy leads to an increase in nuclear emission, possibly
moving the nuclear flux above the limiting magnitude, allowing one to detect
intrinsically fainter Type 1 quasars. And second, the host galaxy dilutes the
AGN emission, making it harder to identify whether the galaxy nucleus hosts
a Type 1 quasar. The exact details of these effects will depend on
the distribution of Type 1 quasar host galaxy luminosities and morphologies
at fixed $M_{BH}$, as well as the Type 1 quasar identification algorithm of a
particular survey. As such, it is unclear how the host galaxy would affect
the $M_{BH}$ completeness limits, and our estimated limits for $i < 24$
should merely be viewed as suggestive.

\section{The Eddington Ratio Distribution}
\label{s-bherf}

\subsection{Number Densities}
\label{s-erat_ndens}

\begin{figure*}
  \begin{center}
    \scalebox{0.7}{\rotatebox{90}{\plotone{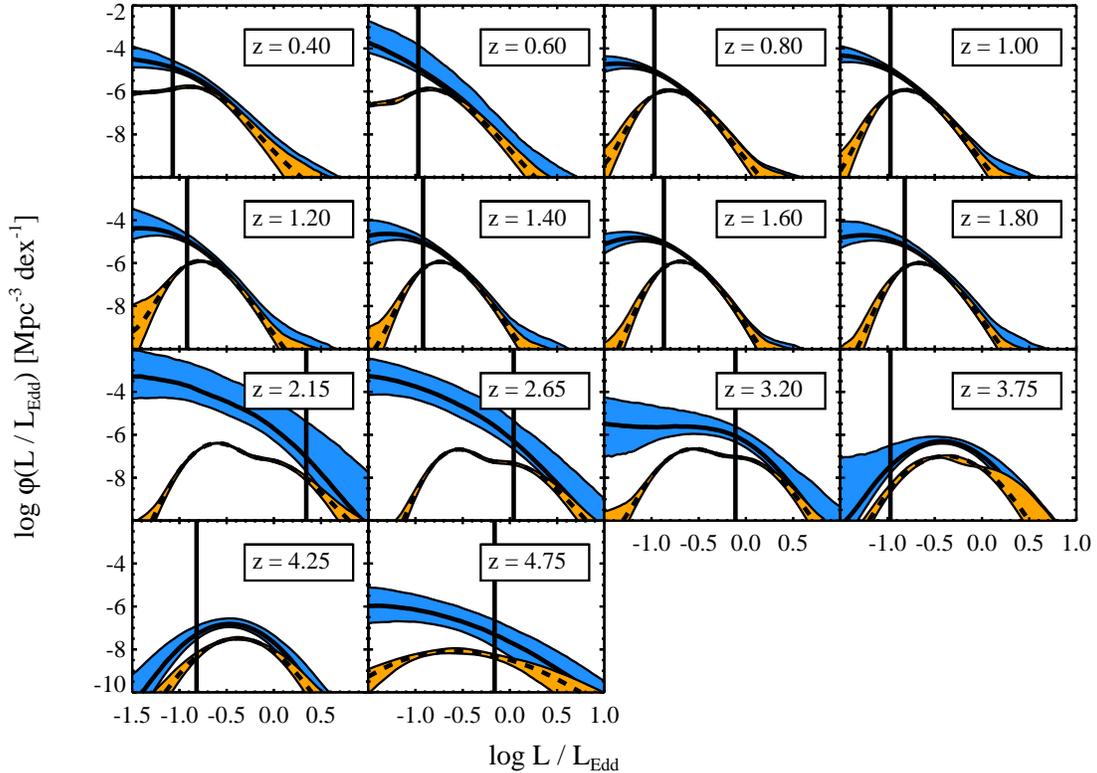}}}
    \caption{The black hole Eddington ratio function for broad line AGN derived from our
      sample. The labeling is the same as in Figure
      \ref{f-bherf}. There is a broad range in $L / L_{Edd}$ for
      Type 1 quasars, and there is no evidence for a turn-over in the BHERF
      down to $L / L_{Edd} \sim 0.07$, except for possibly in the
      redshift range $3.20 \lesssim z \lesssim 4.75$. \label{f-bherf}}
  \end{center}
\end{figure*}

Our estimated Type 1 quasar black hole Eddington ratio function (BHERF) is shown
in Figure \ref{f-bherf} and reported in Table \ref{t-bherf}. Also shown is
the flux-limited BHERF and the value of $L / L_{Edd}$ below which our sample
becomes $< 10\%$ complete. As in Paper I, the BHERF was calculated from the model BHMF and
luminosity function assuming a bolometric correction to $\nu
L_{\nu}(2500$\AA$)$ of $C_{2500} = 5$.
%\textbf{[We used $C_{2500}=5$ in Paper
%I. Should we be consistent here and use the same value?]}, which was obtained
%by scaling a bolometeric correction of 9 at 5100\AA\
%\citep{kaspi2000,vasud07} using the mean spectral slope \textbf{[specify the
%average value of $\mathbf{\alpha_\nu=???}$ that we use here]} of our sample.
In general, the SDSS quasar sample is incomplete at $L / L_{Edd} \lesssim
0.07$. The comoving number densities of SMBHs in Type 1 quasars increase
toward lower Eddington ratio, a trend which continues beyond the
incompleteness limit. The only exceptions are the $z = 3.75$ and $z = 4.25$
bins, which display evidence for a peak in the BHERF at $L / L_{Edd} \approx
0.3$. However, the BHERFs in these bins are derived from the \CIV\ line and
may be subject to systematics resulting from an outflowing wind component
\citep[e.g.,][]{shen08,richards2011}, so it is unclear if this peak
is real. Indeed, the redshift bins before and after these two do not show any
evidence for a peak in the BHERF, although they also employ the \CIV\
line. In addition, we note that because we adopted a more flexible
model for the Eddington ratio distribution, our constraint on the
redshift evolution of the {\em mean} Eddington ratio is much poorer
than that in Paper I, but is generally consistent with the trend found
in Paper I.

\begin{deluxetable}{lcccc}
\tabletypesize{\scriptsize}
\tablewidth{0pt}
\tablecaption{Type 1 Quasar Black Hole Eddington Ratio Functions \label{t-bherf}}
\tablehead{$\bar{z}$
& $\log L / L_{Edd}$
& $\log \Phi_{-}$\tablenotemark{a}
& $\log \Phi_0$\tablenotemark{b}
& $\log \Phi_{+}$\tablenotemark{c} \\
&
& \multicolumn{3}{c}{(Mpc$^{-3}{\rm dex}^{-1}$)}
}
\startdata
0.400 & -1.50 & -4.83 & -4.45 & -3.89  \\
0.400 & -1.40 & -4.84 & -4.52 & -4.00  \\
0.400 & -1.30 & -4.86 & -4.60 & -4.15  \\
0.400 & -1.20 & -4.91 & -4.69 & -4.31  \\
0.400 & -1.10 & -5.00 & -4.83 & -4.48  \\
0.400 & -1.00 & -5.13 & -4.98 & -4.68  \\
0.400 & -0.90 & -5.31 & -5.18 & -4.91  \\
0.400 & -0.80 & -5.54 & -5.41 & -5.17  \\
0.400 & -0.70 & -5.83 & -5.70 & -5.45  \\
0.400 & -0.60 & -6.17 & -6.02 & -5.76  \\
0.400 & -0.50 & -6.57 & -6.40 & -6.11  \\
0.400 & -0.40 & -7.03 & -6.81 & -6.49  \\
0.400 & -0.30 & -7.54 & -7.25 & -6.90  \\
0.400 & -0.20 & -8.09 & -7.72 & -7.31  \\
0.400 & -0.10 & -8.65 & -8.18 & -7.70  \\
0.400 & 0.000 & -9.22 & -8.63 & -8.08  \\
0.400 & 0.100 & -9.79 & -9.08 & -8.46  \\
0.400 & 0.200 & -10.3 & -9.49 & -8.77  \\
0.400 & 0.300 & -10.8 & -9.88 & -9.07  \\
0.400 & 0.400 & -11.3 & -10.2 & -9.37  \\
0.400 & 0.500 & -11.9 & -10.6 & -9.61  \\
0.400 & 0.600 & -12.5 & -10.9 & -9.85  \\
0.400 & 0.700 & -13.1 & -11.3 & -10.0  \\
0.400 & 0.800 & -13.6 & -11.6 & -10.3  \\
0.400 & 0.900 & -14.1 & -12.0 & -10.5  \\
0.400 & 1.000 & -14.6 & -12.3 & -10.7
\enddata

\tablenotetext{a}{Lower boundary on the region containing $68\%$ of the
  posterior probability for the BHERF, i.e., the $16^{\rm th}\%$
  percentile of the posterior distribution.}
\tablenotetext{b}{Posterior median for the BHERF.}
\tablenotetext{c}{Upper boundary on the region containing $68\%$ of the
  posterior probability for the BHERF, i.e., the $84^{\rm th}\%$
  percentile of the posterior distribution.}

\tablecomments{The full table is available in the electronic version of the
paper. Tabulated here are the results for the $\beta=0$ error model.
Results for the $\beta\neq 0$ error model are similar to those presented in
Paper I.}
\end{deluxetable}

\begin{figure}
  \begin{center}
    \scalebox{0.9}{\rotatebox{90}{\plotone{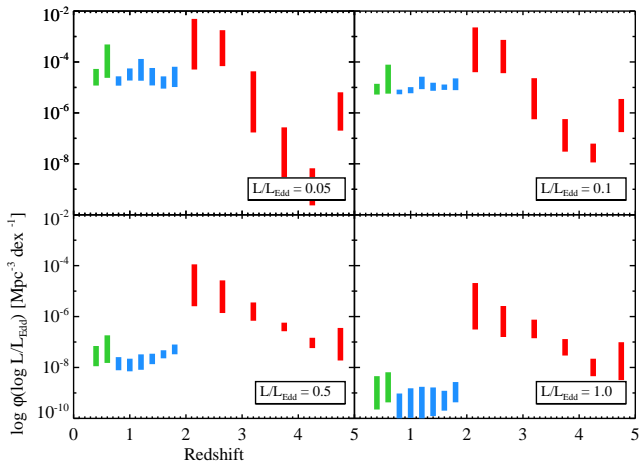}}}
    \caption{Evolution in the comoving number densities of Type 1 quasars at
      four different values of $L / L_{Edd}$. The labeling is the same as
      in Figure \ref{f-bhmf_evol}. The number densities of Type 1 quasars are
      fairly constant in redshift below $z < 2$ for all Eddington
      ratio bins. However, the number densities of Type 1 quasars radiating at
      lower values of $L / L_{Edd}$ drop off more steeply toward
      higher redshift at $z > 2$. \label{f-bherf_evol}}
  \end{center}
\end{figure}

Figure \ref{f-bherf_evol} shows the evolution in the comoving number
densities at four different values of $L / L_{Edd}$. The number densities at
$L / L_{Edd} \gtrsim 0.05$ are roughly constant at $z < 2$, implying that
there is no `downsizing' in Eddington ratio at these redshifts. However, the
number densities of Type 1 quasars radiating at $L / L_{Edd} \lesssim 0.1$
decline rapidly at $z \gtrsim 3$ while the number densities of Type 1 quasars
radiating at $L / L_{Edd} \gtrsim 0.5$ are similar at $z \lesssim 2$ and $z
\gtrsim 3$. This may be evidence for downsizing in Eddington ratio, in which
the number densities of Type 1 quasars at low Eddington ratios increases
rapidly from $z \sim 4$--$5$ to $z \sim 3$. However, we caution that this
trend is primarily driven by the $z = 3.75$ and $z = 4.25$ redshift
bins, which are the only ones that show a peak in the BHERF above the
SDSS completeness limit.

There is an apparent discontinuity in the number densities across $z \sim 2$,
with the number densities at $z \sim 2.15$ being $\sim 1$--$2$ orders of
magnitude larger than that at $z \sim 1.8$. While the uncertainties on the
number densities at $2 < z < 3$ are large, making this only a $1$--$2\sigma$
effect, there are a couple of other issues with this redshift range that are
worth commenting on. For one, $z \sim 2$ marks the transition between mass
estimates calculated from \MgII\ and \CIV, so there may be systematic
differences among these lines. However, we do not see the same effect in the
evolution of the BHMF, and the number densities of the virial mass estimates
(i.e., the binned BHMF) typically show continuity between the redshift bins
where the mass estimates switch emission lines (Paper I), suggesting that the
use of different emission lines is not the dominant reason for this
discontinuity. Instead, the apparent discontinuity in number density may be
primarily due to systematic errors in our incompleteness correction. The
three redshift bins $z = 2.15, 2.65,$ and $3.20$ correspond to redshifts
where quasars colors are similar to star colors, making the SDSS quasar color
selection incomplete at these redshifts. The completeness can be as low as
$\sim 5\%$ in this redshift range. The color distribution of simulated
quasars does not perfectly match the observed color distribution at these
redshift \citep{richards2006}, suggesting that there may be systematic
uncertainties in the estimated selection function. In addition, the
completeness of the selection algorithm at these redshifts depends on the
quasar optical/UV SED, which in turn has been found to depend on $L /
L_{Edd}$ \citep[e.g.,][]{bonning2007}. However, any dependence of this
completeness on $L / L_{Edd}$ is not accounted for in the SDSS selection
function, introducing systematic error if the quasar colors do depend on $L /
L_{Edd}$. Similarly, the Type 1 quasar bolometric correction also depends on
$L / L_{Edd}$ and $M_{BH}$ \citep[e.g.,][]{vasud07,kelly08,vasud09}. Our use
of a constant bolometric correction may exasperate the systematics over this
redshift range. Because the estimated number densities are unstable in bins
of $L / L_{Edd}$ that are significantly incomplete (see \S\ref{s-mbh_ndens}),
there is likely a significant additional systematic uncertainty that is not
reflected in the error bars on the number densities at $2 \lesssim z \lesssim
3.2$. Therefore, we do not consider the apparent discontinuity in the number
densities at $z \sim 2.15$ to be real.

\begin{figure*}
  \begin{center}
    \scalebox{0.7}{\rotatebox{90}{\plotone{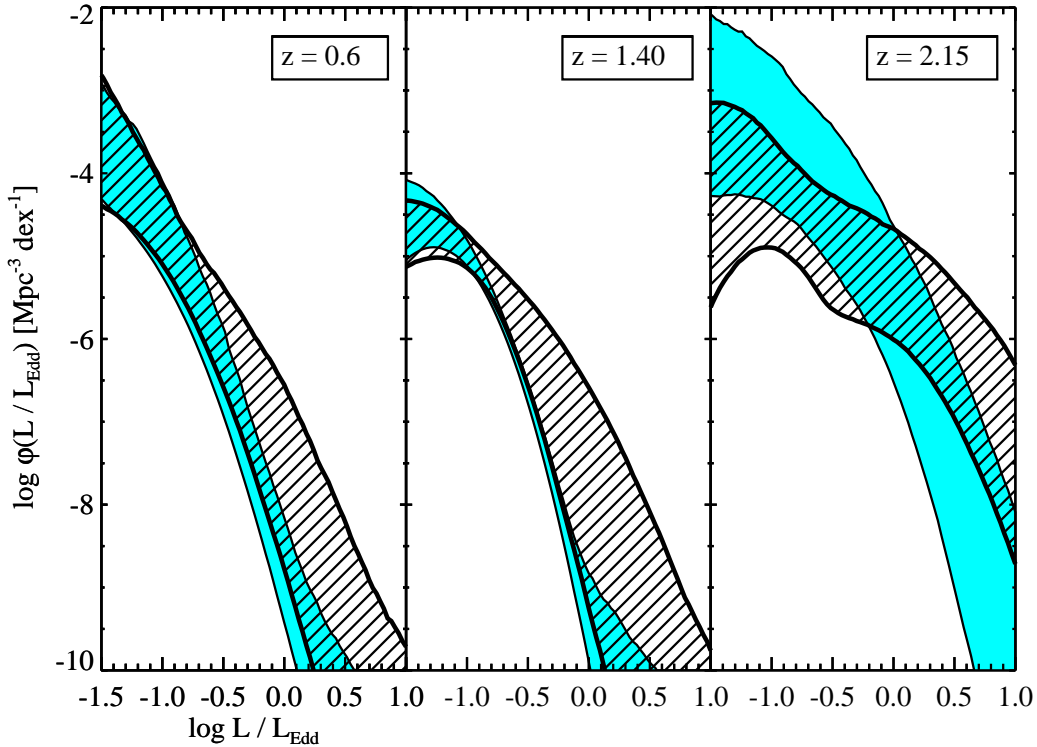}}}
    \caption{Same as Figure \ref{f-bhmf_compare}, but for the Eddington
      ratio function. Unlike for the BHMF, there is a more significant
      difference in the BHERFs derived under the two models for the
      virial mass estimate errors. In all cases the uncertainties on the
      BHERF are larger for the mode that includes a luminosity-dependent
      bias. In addition, the BHERF under the model with $\beta \neq 0$
      is shifted toward higher values $L / L_{Edd}$. This shift
      is small in the redshift bins using \hbeta, but increases to
      $\approx 0.5$ dex for the redshift bins using \MgII\ and \CIV.}
    \label{f-bherf_compare}
  \end{center}
\end{figure*}

Unlike with the BHMF, the BHERF is noticeably different when we allow for a
luminosity-dependent bias. In Figure \ref{f-bherf_compare} we compare the
BHERF derived from the models with and without a luminosity-dependent bias
for three representative redshift bins. This difference is negligible for the
\hbeta-based mass estimates, for which $\beta \approx 0$. However, in the
redshift bins where \MgII\ and \CIV\ mass estimates are used the model that
includes the luminosity-dependent bias leads to BHERFs which are more
uncertain and fall off flatter toward higher $L / L_{Edd}$, implying a larger
number of Type 1 quasars radiating near the Eddington limit. In addition,
while the number densities at the high $L / L_{Edd}$ end of the BHERF depend
on the value of $\beta$, the evolution in the number densities at fixed $L /
L_{Edd}$ is not as affected by the value of $\beta$. The evolution results
obtained for the luminosity-dependent bias model are not significantly
different from those obtained for the $\beta = 0$ model.

\subsection{Distribution of Eddington Ratio as a Function of Black
  Hole Mass}
\label{s-mlplane}

\begin{figure*}
  \begin{center}
    \scalebox{0.7}{\rotatebox{90}{\plotone{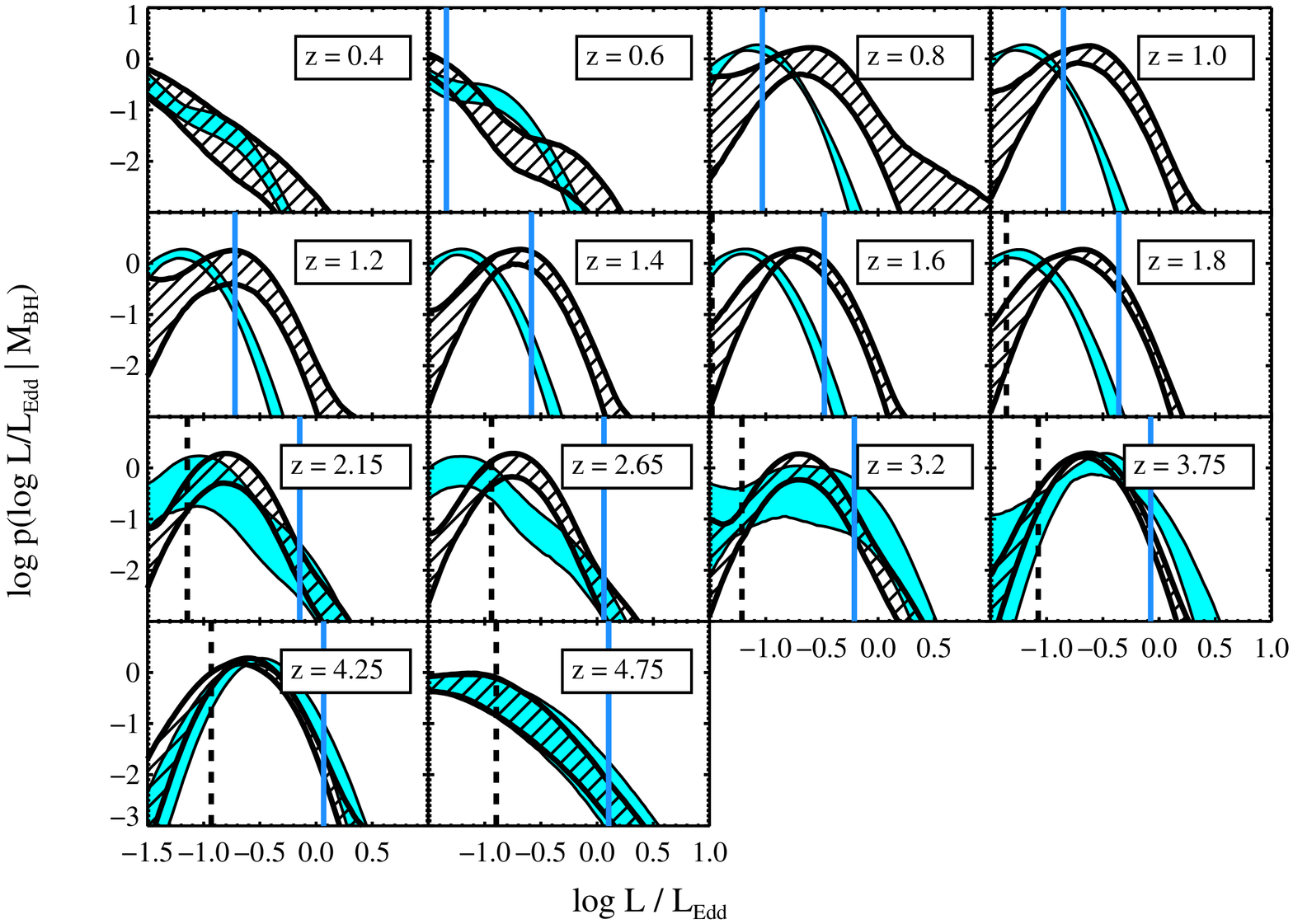}}}
    \caption{Conditional probability distributions for Eddington ratio
      at $M_{BH} = 5 \times 10^8 M_{\odot}$ (cyan region) and $M_{BH}
      = 5\times 10^9 M_{\odot}$ (diagonal black line filled region); both
      regions contain $68\%$ of the posterior probability in $p(\log L
      / L_{Edd} | M_{BH})$ as a function of $M_{BH}$. The solid blue
      vertical line marks the incompleteness limit in $L / L_{Edd}$
      for $M_{BH} = 5 \times 10^8 M_{\odot}$, while the vertical
      dashed black line marks the limit for $M_{BH} = 5 \times 10^9
      M_{\odot}$. Note that for $M_{BH} = 5 \times 10^9 M_{\odot}$ the
      limit occurs below $\log L / L_{Edd} = -1.5$ at $z \lesssim
      1.6$. \label{f-cprob_erat}}
  \end{center}
\end{figure*}

In Figure \ref{f-cprob_erat} we show the conditional probability
distribution of $\log L / L_{Edd}$ at $M_{BH} = 5 \times 10^8
M_{\odot}$ and $M_{BH} = 5 \times 10^9 M_{\odot}$ for $z = (0.6, 1.6, 2.65)$; note that these
quantities integrate to one and are not the same as the BHMF and
BHERF. We show $p(\log L / L_{Edd}|M_{BH})$ for both the model which
assumes $\beta = 0$ and the model with free $\beta$. As with the
BHERF, the estimated conditional Eddington ratio distributions are
broader and more uncertain, especially for the redshifts where the
\MgII\ line is used.

In general the conditional Eddington ratio distributions mimic the
behavior seen in Figure \ref{f-cprob_erat}, with 
$p(\log L / L_{Edd}|M_{BH})$ being similar for $M_{BH} = 5 \times
10^8 M_{\odot}$ and $M_{BH} = 5 \times 10^9 M_{\odot}$ at $z \sim
0.6$, but with the distribution of $L / L_{Edd}$ being shifted to
larger values for $M_{BH} = 5 \times 10^9 M_{\odot}$ compared to
$M_{BH} = 5 \times 10^8 M_{\odot}$. In addition, at higher redshifts
the distributions of $\log L / L_{Edd}|M_{BH}$ exhibit a peak around
$L / L_{Edd} \sim 0.1$. The peak in $p( \log L / L_{Edd}|M_{BH})$ is below the
SDSS completeness limit for $M_{BH} \sim 5 \times 10^8 M_{\odot}$ at
$z \gtrsim 1.5$, so it is unclear if this is a real feature. On the
other hand, for $M_{BH} \sim 5 \times 10^9 M_{\odot}$ this peak does occur above
the completeness limit.

% In Figure \ref{f-mlplane} we show the median and $95^{\rm th}$ percentile of the
% distribution of Type 1 quasar Eddington ratio at fixed $M_{BH}$; note
% that the $95^{\rm th}$ percentile is the value of $L / L_{Edd}|M_{BH}$
% at which $95$\% of quasars are below this value. At most
% redshifts the median value of $L / L_{Edd}|M_{BH}$ is below our
% incompleteness limits, so we limit our discussion to trends involving
% the $95^{\rm th}$ percentile of the $L / L_{Edd}|M_{BH}$ distribution;
% the median values are simply shown for reference and in general exhibit
% similar trends as the $95^{\rm th}$ percentile. 

At both high ($z \gtrsim 3.2$) and low ($z \lesssim 0.6$) redshifts
the $L / L_{Edd}$ distribution is relatively independent of
$M_{BH}$. However, at redshifts $0.8 \lesssim z \lesssim 2.65$ the
distribution of $L / L_{Edd}$ at fixed $M_{BH}$ shifts to larger
Eddington ratios from $M_{BH} \sim 5 \times 10^8 M_{\odot}$ to $M_{BH}
\sim 5 \times 10^9 M_{\odot}$. This therefore implies that at $0.8
\lesssim z \lesssim 2.65$ Type 1 quasars with more massive black holes
are more likely to be radiating near the Eddington limit. However, we
note that this redshift range is dominated by mass estimates derived
from \MgII, and it is possible that systematic effects with this line
may be driving some of these results. The results in the last two
redshift bins in this range are derived from the \CIV\ line mass
estimates, and although they do show a shift in the distribution
toward larger $L / L_{Edd}$ for larger $M_{BH}$ the difference in the
Eddington ratio distributions at $M_{BH} = 5 \times 10^8 M_{\odot}$
and $M_{BH} = 5 \times 10^9 M_{\odot}$ are not as strong as that seen
in the \MgII\ bins. In fact, in the two $z < 3$ bins that use the
\CIV\ estimate ($z = 2.15$ and $z = 2.65$), the fraction of Type 1
quasars radiating near the Eddington limit is the same for both mass
bins within the uncertainties. The results from this section are
similar for the model that includes a luminosity-dependent bias, with
the exceptions that the distributions of $L/L_{Edd}|M_{BH}$ were
broader and more uncertain.

These results are in contrast to the conclusions reached by
\citet{steinhardt2010a}. These authors used virial mass estimates from the
SDSS DR5 quasar sample of \citet{shen08} to argue that the most massive Type
1 quasars fall short of the Eddington limit, forming what they called a
`sub-Eddington boundary' \citep[but see][]{rafiee2011a}. \citet{steinhardt2010a} quantified this trend
through the $95^{\rm th}$ percentile of luminosities above the peak in the
distribution of estimated bolometeric luminosities for the sample, finding
that the $95^{\rm th}$ percentile in luminosity increased slower with
$M_{BH}$ than would be expected from a constant value of $L / L_{Edd}$. We
consider the likely reason for their different conclusion to be uncorrected
incompleteness. While it is true that for a flux-limited sample the $95^{\rm
th}$ percentile in $L / L_{Edd}$ at fixed $M_{BH}$ mass does increase with
decreasing $M_{BH}$, this is not necessarily true of the Type 1 quasar
population as a whole. The reason for this is because the $95^{\rm th}$
percentile in $L / L_{Edd}$ at fixed $M_{BH}$ for a flux-limited sample is an
overestimate of the true $95^{\rm th}$ percentile due to the loss of the
faint end of the population. For smaller values of $M_{BH}$ one loses a
larger fraction of the lower $L / L_{Edd}$ part of the population, increasing
the bias in the $95^{\rm th}$ percentile inferred from the distribution of
Type 1 quasars that are actually bright enough to be detected. Indeed, for
the least massive black holes in a flux-limited sample one cannot even detect
the $95^{\rm th}$ percentile of the Eddington ratio distribution, as it falls
below the flux limit. This leads to a spurious increase in the $95^{\rm th}$
percentile of the $L / L_{Edd}$ distribution with decreasing $M_{BH}$.
In addition, the errors of virial BH mass estimates stretch the
distribution in the mass-luminosity plane along the mass direction, causing
artificial flattening and an apparent ``tilt'' away from the Eddington limit
towards higher virial BH masses (see Fig.\ 8 of Paper I), which could be
incorrectly recognized as visual evidence for a sub-Eddington
boundary. However, we note that this effect is not as strong as the bias caused by
incompleteness.

In this work we have corrected for incompleteness in flux-limited
samples and errors in virial BH mass estimates, assuming our statistical
model and the SDSS selection function, and thus attempt to recover the true
intrinsic trends in the high $L / L_{Edd}$ tail with $M_{BH}$. Upon doing so,
we do not find any evidence that higher mass Type 1 quasars are less likely
to be observed at high $L / L_{Edd}$, but in fact at certain redshifts may be
more likely to have high $L / L_{Edd}$. Therefore, the observed dirth of Type
1 quasars having both high $M_{BH}$ and high $L / L_{Edd}$ is not caused by a
change in the shape of the tail of the Eddington ratio distribution, but is
instead caused by the rapidly decreasing number densities of Type 1 quasars
toward higher masses. Because the most massive SMBHs in Type 1 quasars are
rare by definition, we would not expect those that we do observe to also occupy
the rare high-$L / L_{Edd}$ region of the mass-luminosity plane.

At low redshift past the peak of quasar activity, the most massive
SMBHs are probably accreting mostly at very low Eddington ratios, but
are no longer shining in the form of Type 1 quasars.  We do not find
any statistical evidence of a sub-Eddington boundary in the
mass-luminosity plane of type 1 quasars, and earlier claims of such a
boundary are likely the result of uncorrected incompleteness.

\subsection{Do quasars obey the Eddington Limit?}
\label{s-elimit}

\begin{figure}
  \begin{center}
    \scalebox{0.9}{\rotatebox{90}{\plotone{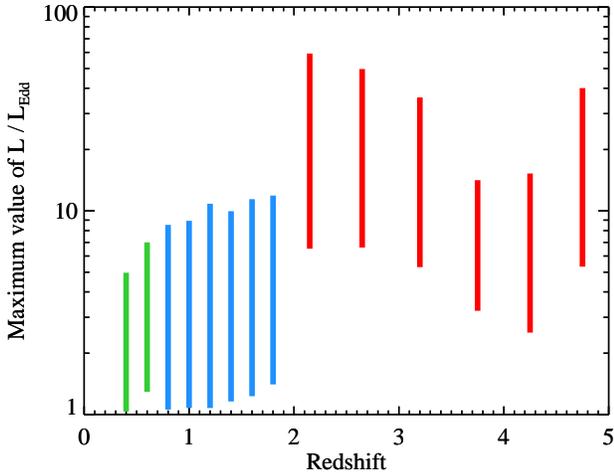}}}
    \caption{Constraints on the maximum Eddington ratio that could be observed
      in a broad-line quasar as a function of redshift. Symbols are
      as in Figure \ref{f-bhmf_evol}. There is no evidence for
      significantly super Eddington radiation, and the maximum value of $L / L_{Edd}$
      for quasars implied by our BHERF is $L / L_{Edd} \sim 3$. \label{f-eddlimit}}
  \end{center}
\end{figure}

Similar to the calculation of the maximum mass performed in
\S~\ref{s-most_massive}, we can also calculate the probability distribution
of the highest value of $L / L_{Edd}$ from a population of Type 1 quasars
drawn from our estimated BHERF. The maximum value of $L / L_{Edd}$ as a
function of redshift is shown in Figure \ref{f-eddlimit}. Our BHERF implies
that the maximum value of $L / L_{Edd}$ for a Type 1 quasar is $L / L_{Edd}
\sim 3$, although values between $L / L_{Edd} = 1$ and $L / L_{Edd} = 10$ are
well within the uncertainties. The results from using the model with a
luminosity-dependent bias suggests values of the maximum Eddington ratio a
factor of $\sim 2$ higher, which is smaller than the uncertainties within
individual redshift bins. The maximum values of $L / L_{Edd}$ are somewhat
divergent at $2 < z < 3.2$, having values near $L / L_{Edd} \sim 10$, but, as
discussed above, this range is subject to large statistical and
systematic uncertainties. In addition, our results suggest that quasars at
$z \sim 4$ may obtain a maximum value of $L / L_{Edd}$ that is
slightly larger than quasars at $z \lesssim 2$, but the error bars are large.
Considering this, our results are consistent with a maximum value of $L /
L_{Edd} \sim 3$ across all redshift bins. In addition, there is additional
systematic uncertainty in the maximum value of $L / L_{Edd}$ caused by our
assumption of a constant bolometeric correction. Taking this into account, we
do not find any evidence that Type 1 quasars violate the Eddington limit by
more than a factor of a few.

\subsection{Completeness in Eddington Ratio}
\label{s-erat_completeness}

\begin{figure}
  \begin{center}
    \scalebox{0.9}{\rotatebox{90}{\plotone{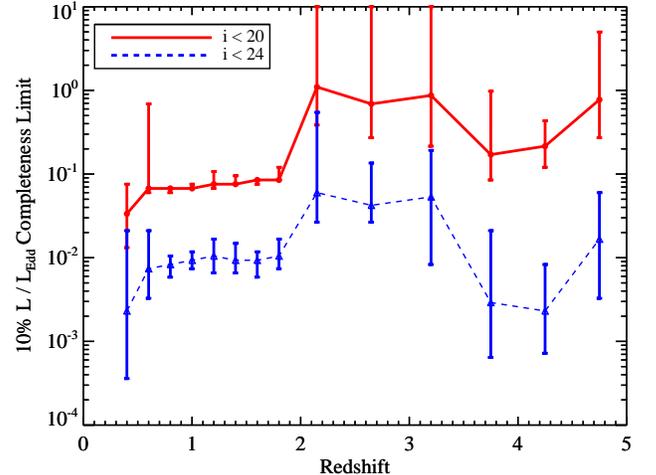}}}
    \caption{The values of $L / L_{Edd}$ at which a Type 1 quasar survey
      becomes only $10\%$ complete under a limiting magnitude of $i < 20$
      and $i < 24$. The labeling is the same as in Figure
      \ref{f-mass_completeness}. The SDSS becomes highly incomplete below $L
      / L_{Edd} \sim 0.07$. \label{f-erat_completeness}}
  \end{center}
\end{figure}

Similar to the calculation performed in \S~\ref{s-mass_completeness}, in
Figure \ref{f-erat_completeness} we show the Eddington ratio at which a
survey becomes only $10\%$ complete as a function of redshift for two
limiting $i$-band magnitudes. The results are similar for the model with a
luminosity-dependent bias. Surveys with a limiting magnitude of $i = 20$ are
largely incomplete at $L / L_{Edd} \lesssim 0.07$, while surveys with a
limiting magnitude of $i = 24$ start to become severely incomplete at $L /
L_{Edd} \lesssim 0.01$. The increase in the completeness limits at $2 < z <
3.2$ mirror other anomalous trends in $L / L_{Edd}$, and the uncertainties
are large. As discussed earlier in this work, the SDSS color selection
algorithm has difficulty distinguishing quasars from stars in this redshift
range, and this may introduce systematic trends with $L / L_{Edd}$.
Therefore, we do not consider the `bump' in the completeness limit at $2 < z
< 3.2$ to be real. These derived completeness limits are broadly consistent
with the lack of Type 1 quasars having $L / L_{Edd} < 0.01$ observed in
\emph{COSMOS} by \citet{trump2011}, whose sample starts to become incomplete
at $i > 23$.

\section{Discussion}
\label{s-discussion}

The results that we have obtained in this work with regard to AGN
demographics provide an important window into SMBH growth and AGN fueling,
especially when interpreted within the greater context of earlier
observational and theoretical work on AGN number densities, clustering, and
host galaxy properties. In particular, a picture is emerging where the
fueling mechanisms of SMBH growth are complex and varied. In this section we
discuss an interpretation of our results that incorporates other recent
observational results; many of the ideas discussed here are treated in
greater detail in various theoretical papers
\citep[e.g.,][]{stochacc,hopkins_long,croton06,bower06,monaco2007,hopkins2008,hh09,fanidakis2012}.

\subsection{Black Hole Growth Time Scales: Evidence for Self-regulated Growth of SMBHs?}
\label{s-selfregulated}

\begin{figure}
  \begin{center}
    \scalebox{0.9}{\rotatebox{90}{\plotone{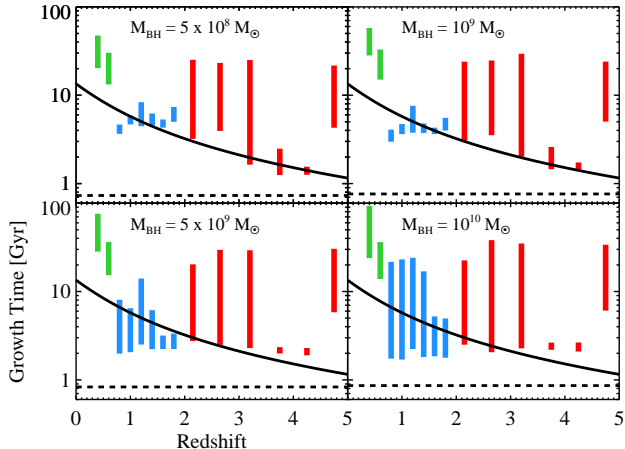}}}
    \caption{Estimated typical time needed for a black hole to grow from a seed mass of
      $M_{BH}^{\rm seed} = 10^6 M_{\odot}$ to four different final
      masses as a function of redshift, assuming a radiative
      efficiency of $\epsilon_r = 0.1$ and the typical Eddington ratio
      for that mass and redshift bin. The solid line marks the
      age of the universe as a function of $z$, the dashed line marks
      the growth time assuming $L / L_{Edd} = 1$, and the rest of the
      labeling is the same as that in Figure
      \ref{f-bhmf_evol}. Although the uncertainties are large, the
      growth times for SMBHs in Type 1 quasars having $M_{BH} \gtrsim 5 \times
      10^8 M_{\odot}$ are comparable to or greater than the age of the
      universe, suggesting an earlier phase of accelerated obscured
      growth. The growth times at $z \lesssim 0.6$ are also longer
      than the age of the universe, possibly reflecting a shift to
      fueling by mass loss from evolved stellar populations.
      \label{f-growth_times}}
  \end{center}
\end{figure}

In Figure \ref{f-growth_times} we show the estimated typical growth time as a function
of redshift needed for SMBHs in Type 1 quasars to obtain masses of $M_{BH} / M_{\odot} = 5 \times
10^8, 10^9, 5 \times 10^9,$ and $10^{10}$. We estimate the time needed for a SMBH
to grow from a seed mass of $M_{\rm seed}$ to a observed mass of
$M_{BH}$ at a redshift $z$ as
\begin{equation}
t_{BH} = \frac{t_s}{10\epsilon_r E(L / L_{Edd}|M_{BH},z)} \ln \frac{M_{BH}}{M_{\rm seed}}
\label{eq-growth_time}
\end{equation}
where $t_s = 4.3 \times 10^7\ {\rm yr}$ is the Salpeter ($e$-folding) time scale assuming a
radiative efficiency of $\epsilon_r = 0.1$ and $E(L /
L_{Edd}|M_{BH},z)$ is the mean value of $L / L_{Edd}$ at fixed
$M_{BH}$ and $z$. We calculated growth time scales assuming a seed
mass of $M_{\rm Seed} = 10^6 M_{\odot}$ and a radiative efficiency of $\epsilon_r = 0.1$
\citep[e.g.,][]{yu02,davis2011}. The assumption in these calculations is that the SMBHs
in each mass and redshift bin accrete at a time-averaged rate relative to Eddington that
is equal to the population average of Type 1 quasars in that
bin. While this is not true for every Type 1 quasar in each bin, it
should be a good approximation for a representative quasar in each
bin; hence the association of these time scales with the `typical'
growth time for a SMBH in 
that bin. In addition, these calculations implicitly assume a duty cycle of
unity. For comparison, we also show the age of the universe as a function
of $z$. The results were similar for the model with a luminosity-dependent
bias, but the error bars on the growth times were larger.  The value of $\epsilon_r = 0.1$
corresponds to the radiative efficiency of a moderately spinning black hole,
and the value of $M_{BH} = 10^6 M_{\odot}$ for the seed black holes
corresponds to the largest values predicted by models for SMBH seed formation
\citep[for a review, see][]{volonteri2010}. In particular, SMBH seed masses
of $M_{BH} = 10^6 M_{\odot}$ represent the high-end of the distribution
predicted from models where SMBH seeds form from the direct collapse of gas
\citep[e.g.,][]{vol08,vol10,natarajan2012}

For many of the mass and redshift bins at $z \gtrsim 2$ the growth times are
comparable to or longer than the age of the universe, although the
uncertainties are large for several of the bins. If the remnants of
Pop III stars are the SMBH seeds then values of the seed masses are $M_{BH}
\sim 100 M_{\odot}$ \citep[e.g.,][]{fryer2001,madau2001} and the growth times
will be a factor of $\sim 2$ longer. Similarly, if the black holes are
rapidly spinning, as expected from continuous accretion in a disk geometry,
then their growth time will be a factor of $\sim 3$ longer due to the
increased radiative efficiency. In addition, if the duty cycles of
Type 1 quasar activity are less than unity than the growth time will
be longer by a factor equal to the inverse of the duty cycle. On the
other hand, we have neglected mergers of
black holes in these growth times, which may make a non-negligible
contribution to shortening the growth times
\citep[e.g.,][]{sesana2007,tan09}. Similarly, if most quasars at high $z$ host
SMBHs that are not spinning the radiative efficiency is $\epsilon_r <
0.1$ and the growth times will be shorter than what we estimate.

Because our estimated typical SMBH growth time scales are comparable
to, or longer than, the age of the universe, they imply that many
SMBHs underwent an earlier phase of accelerated growth (e.g., with
higher Eddington ratio) when we would not have observed them as Type 1
quasars. Similar conclusions have been reached by \citet{netzer07a},
K10, and \citet{trakh2011}. It is possible that we do not observe
these early rapidly growing SMBHs as Type 1 quasars because they are
obscured. Alternatively, it is also possible that the typical Eddington
ratio is near unity for Type 1 quasars with masses representative of
an earlier phase in their growth (e.g., $M_{BH} \lesssim 5 \times 10^8
M_{\odot}$). In this case, we would have missed these rapidly accreting Type 1 quasars
because they would have had smaller masses, and therefore would have fallen below our
completeness limit. Both possibilities exist but cannot be
distinguished by our data.

%While it is likely that some of the SMBHs at $z \sim 2$
%represent reignition of AGN activity, and thus that the bulk of the growth
%could have happened at earlier times, the typical Eddington ratios of Type 1
%quasars at $2 \lesssim z \lesssim 4.5$ are still too low for all of them to
%have grown during earlier Type 1 quasar phases. Moreover, the number
%densities of Type 1 quasars at $z \sim 2$ are a factor of $\sim 10$--$100$
%larger than those at $z \sim 4$, implying that many of the Type 1 quasars
%observed at $z \sim 4$ are not simply the progenitors of those at $z \sim 2$.

Our results regarding the growth times of typical SMBHs in Type 1 quasars is
in agreement with expectations from models for self-regulated black hole
growth\citep[e.g.,][]{sanders1988,Sanders_Mirabel_1996,hopkins_long}.
Within the framework of these models, the black hole is enshrouded in
material and undergoes Eddington limited obscured growth. The broad line
quasar phase begins after the obscuring material disappears. This can happen
because the black hole grows to the point that feedback energy from it
becomes powerful enough to `blow' the obscuring material away
\citep[e.g.,][]{quasar_evol,hopkins_long}. Alternatively, the obscuring
material may disappear because it becomes consumed by star-formation in the
bulge. The Type 1 quasar phase is expected to persist until the accretion
rate drops low enough to switch to a radiatively inefficient accretion flow
\citep[e.g.,][]{churazov05,Cao_2005,shen07b,hhqh09}. We note that
during the blow-out phase there may still be geometry-dependent obscuration
caused by a dusty torus, as invoked by the traditional AGN unification models
\citep{antonucci1993,urry1995}. Our results are therefore consistent with a
variety of models for growing the most massive SMBHs at $z \gtrsim 2$, so
long as SMBHs exhibited enhanced accretion rates relative to Eddington below $M_{BH} \sim 5 \times
10^{8} M_{\odot}$.

At redshifts $z \lesssim 0.8$ our calculated black hole growth times are a
factor of $\sim 2$--$3$ longer than the age of the universe for SMBHs with
$M_{BH} > 5 \times 10^8 M_{\odot}$, reflecting the smaller values of $L /
L_{Edd}$ at these redshifts. Similar results were obtained by
\citet{heckman2004} and \citet{kauffmann2009} using a sample of low-$z$ Type
2 AGN. These low-$z$ AGN likely experienced an earlier stage of rapid growth,
and instead we are currently witnessing a reignition of weaker AGN activity
reflected in their lower accretion rates and longer growth times. As we discuss below in \S~\ref{s-fueling}, the
Type 1 quasars probed by our study most likely live in elliptical
galaxies, and therefore this low-order AGN activity is probably fueled by stellar mass loss from
evolved stars. Although there is little overlap in $M_{BH}$ between the
massive systems in our sample and the less massive ones in the sample of
\citet{kauffmann2009}, at $M_{BH} \sim 3 \times 10^8 M_{\odot}$ they derive
an Eddington ratio distribution which decreases monotonically toward higher
$L / L_{Edd}$, similar to us.

\subsection{Fueling of SMBHs and Triggering of Type 1 Quasar Activity}
\label{s-fueling}

Due to incompleteness, our sample probes the Type 1 quasar population with
$M_{BH} \gtrsim 3 \times 10^8 M_{\odot}$. Locally, SMBHs in this mass range
are predominately found in early-type (i.e., elliptical and S0 galaxies)
galaxies as inferred from the local SMBH BHMF \citep{yu04,yu08}. This
therefore suggests that our study is dominated by SMBHs that currently live
in elliptical and S0 host galaxies. Indeed, many studies find that the host
galaxies of optically-identified AGN are dominated by early type galaxies
\citep[e.g.,][]{kauffmann2003,dunlop2003,zakamska2006,kot2007}. Currently, the favored
mechanism for creating elliptical galaxies is through galaxy mergers
\citep[e.g.,][]{springel2005,bournaud2005,boylan2006,cox2006}, suggesting
that the Type 1 quasars in our sample have at least in part been fueled by
mergers. This is supported by observations which find evidence of past
interactions in many local elliptical hosts of quasars
\citep[e.g.,][]{bahcall1997,bennert2008}. However, we also note that some
recent theoretical work suggests that spheroid-dominated galaxies may also
form from instabilities in gas-rich disks at high-$z$ \citep{dekel2009};
furthermore, \citet{bournaud2011} has argued that AGN at high-$z$ may form
from similar processes. This said, the fraction of galaxies at high-$z$ that
contain the massive clumps necessary to form spheroids and AGN via secular
processes is potentially low \citep{wuyts2012}, and at best unclear.

As discussed above, the typical growth times that we find for $M_{BH} \gtrsim
5 \times 10^8 M_{\odot}$ SMBHs in Type 1 quasars at $z \gtrsim 2$ are of
order the age of the universe or longer, implying that these SMBHs
experienced an earlier phase of accelerated growth. This is expected
from models for self-regulated SMBH growth. Such
models posit a massive fueling event, such as a major merger, which directs
larger amounts of gas toward the nuclear region, fueling obscured
Eddington-limited SMBH growth. Eventually, the SMBH's growth is quenched
either due to AGN feedback or the consumption of gas due to star formation,
revealing the SMBH as a Type 1 quasar and linking the SMBH's final mass with
properties of the stellar bulge. Indeed, it is likely that most of our Type 1
quasars currently reside in early type galaxies, which show the tightest
$M_{BH}$--$\sigma^*$ relation \citep[e.g.,][]{gult09}. 

Due to incompleteness our sample only probes SMBHs with $M_{BH}
\gtrsim 3 \times 10^8 M_{\odot}$, and it is unlikely that our results
and discussion extend to lower mass SMBHs. Estimates of the local BHMF
for all SMBHs imply that late types should dominate SMBH host galaxies
below $5 \times 10^7 M_{\odot}$
\citep[e.g.,][]{yu04,yu08}. Furthermore, among galaxies that have been
used to define the $M_{BH}$--$\sigma_*$ relationship, late types are
found at $M_{BH} \lesssim 10^8 M_{\odot}$
\citep[e.g.,][]{mcconnell2011}. Many recent studis of X-ray selected
AGN at lower luminosities ($L_{\rm bol}<{\rm a\
few}\times10^{45}\,{\rm erg\,s^{-1}}$) have found that AGN live in
both bulge- and disk-dominated galaxies out to $z \sim 2$, and that their host galaxies
are no more likely to show evidence of disturbances or interaction
signatures than their inactive counterparts \citep[][but see
\citet{koss2010}]{grogin2005,pierce2007,gabor2009,cisternas2011,schawinski2011,povic2012,kocevski2012}. However,
it should be noted that there are difficulties in identifying mergers
as the trigger of AGN activity due to a possible time-lag between the
merger and the initiation of AGN activity
\citep[e.g.,][]{hopkins_faintend,schawinski2010a}, combined with the
rapid fading of the morphological features indicative of a recent
merger \citep{lotz2010}. 

Studies of AGN host galaxy colors have found results consistent with the
morphology studies: namely, that AGN live in both early and late type
galaxies
\citep[e.g.,][]{nandra2007,coil2009,silverman2009,schawinski2009,xue2010,george2011}. The
most recent studies are able to go deeper and find a higher fraction of
AGN in disk galaxies. Indeed, \citet{kocevski2012} found a correlation
between X-ray luminosity and the fraction of AGN in spheroidal
galaxies, and \citet{treister2012} found a correlation between AGN
luminosity and the fraction of host galaxies undergoing a major
merger. These results
suggest that AGN with more massive SMBHs tend to be found in early
types and fueled by major mergers
even out to $z \sim 2$--$3$. However, the larger fraction of AGN with less massive SMBHs in
disk galaxies suggests that these SMBHs are grown through processes
other than a major merger up to the point that we observe them, and it is unclear if they also
experience an earlier phase of accelerated growth. Moreover the
weaker $M_{BH}$--$\sigma_*$ relationship for late type galaxies
\citep[e.g.,][]{graham08,gult09,greene2010a,korm11} suggests that the growth
of these SMBHs is more weakly coupled to the evolution of the host galaxy, if
any relationship exists at all. In contrast, the more massive SMBHs ($M_{BH}
\gtrsim 3 \times 10^8 M_{\odot}$) probed by our Type 1 quasar sample are
grown through a process that also generates a spheroid and places the SMBH on
the local $M_{BH}$--$\sigma_*$ relation, with a major merger of two gas rich
galaxies likely being the dominant fueling mechanism.

\subsection{Downsizing of SMBHs and Duty Cycles of Type 1 Quasar Activity}
\label{s-duty}

In this work we have also found evidence for downsizing in both
$M_{BH}$ and $L / L_{Edd}$. The downsizing in $M_{BH}$ is consistent
with results from SMBHs in Type 1 quasars \citep[e.g.,][ K10, Paper
I]{vest08,labita2009a} and for the entire SMBH population as derived
from the continuity equation methods \citep[e.g.,][]{marc04,merloni04},
and reflects the fact that the most massive SMBHs tended to experience
active phases and the bulk of their growth before less massive
SMBHs. Studies of AGN clustering have concluded that AGN are observed
to reside in dark matter halos of $M_h \sim 3 \times 10^{12}
M_{\odot}$ at all redshifts
\citep[e.g.,][]{porciani2004,croom2005,coil2007,myers2007,shen2007,Shen_etal_2009,Ross_etal_2009,daangela2008,mand2009,hickox2009,hickox2011,white2012}. The
combination of the clustering results and AGN downsizing led
\citet{hickox2009} to argue that AGN activity is triggered when a
SMBH's host halo reaches $M_h \sim 3 \times 10^{12} M_{\odot}$, and
that the observed downsizing in $M_{BH}$ is a reflection of the fact
that the most massive SMBHs in the present epoch are those that
resided in the most massive halos at high redshift, and these are the
halos that reached a critical mass of $M_h \sim 3 \times 10^{12}
M_{\odot}$ first.

We can also use our estimated Type 1 quasar BHMF to estimate the
duty cycle for Type 1 quasar activity. We estimate the duty cycle at
$z \sim 0.4$ and $z \sim 1.0$ for Type 1 quasar activity of SMBHs with
$M_{BH} = 10^9 M_{\odot}$ by taking the ratio of our estimated BHMF to
the local BHMF for all SMBHs at $M_{BH} = 10^9 M_{\odot}$, where the
local BHMF and its uncertainy are taken from the compiliation of
\citet{shank09}. The local BHMF should provide a good estimate of the
BHMF at $z = 0.4$ and $z = 1.0$ as the bulk of the BHMF at $M_{BH}
\sim 10^9 M_{\odot}$ was already in place by $z \sim 1$
\citep[e.g.,][]{merloni08}. We find duty cycles of $6.0^{+3.9}_{-2.4}
\times 10^{-3}$ and $9.1^{+3.8}_{-2.7} \times 10^{-3}$ ($68\%$
credibility intervals) for $M_{BH} =
10^9 M_{\odot}$ at $z = 0.4$ and $z = 1$, respectively. Similarly, we
also estimate the duty cycle of Type 1 quasar activity for $M_{BH} =
10^9 M_{\odot}$ SMBHs at $z = 2$ by comparing with the BHMF for all
SMBHs derived from continuity equations methods. Comparing our active
BHMF with the compilation of $z = 2$ BHMFs shown in Figure 7b of
\citet{bhmf_review}, we estimate the $z= 2$ duty cycle of Type 1
quasar activity to be $> 0.06$. These results show that for SMBHs with
$M_{BH} \sim 10^9 M_{\odot}$ Type 1 quasar activity is a rare and
likely short-lived phenomenon, and was more common at
higher redshift, implying that the massive end of the BHMF was built
up primarily at early times. Similar results were obtained in Paper I
by comparing the cumulative mass densities in Type 1 Quasars with all
SMBHs.

The duty cycles that we estimate are broadly consistent with
those derived from continuity models by \citet{shankar2012} for their
Gaussian + Power-law Eddington ratio model. Moreover, the $z = 1$ duty
cycle that we estimate is very similar to that estimated by K10 from
the SDSS DR3, and our estimated duty cycle at $z = 0.4$ is about an
order of magnitude larger than that at $z < 0.3$ estimated by
\citet{schulze2010}. This latter point implies that the duty cycle of
Type 1 quasar activity for $M_{BH} = 10^9 M_{\odot}$ SMBHs declines
rapidly from $z \sim 0.4$. However, we also note that our $z \sim 0.4$
duty cycle may also be overestimated due to a bias in our
incompleteness correction caused by unaccounted for host galaxy flux
in the SDSS selection function, as discussed in \S~\ref{s-mbh_ndens}
and \S~\ref{s-systematics}. And finally, our estimated growth times at
high redshift imply duty cycles of order unity, as it is difficult to accomodate the
longer growth times required by smaller values of the duty cycle given
that the estimated growth time scales are already comparable to the
age of the universe at these redshifts. Our inferred large duty cycles for luminous
quasars at high redshift are consistent with the large duty cycles
inferred by quasar clustering measurements at $z \gtrsim 3$
\citep[e.g.,][]{white2008,shankar2010,shen2010,bonoli2010}.

\subsection{Eddington Ratio Distributions and Type 1 Quasar Lightcurves}
\label{s-eddrat}

Under self-regulated growth models, during the
Type 1 quasar phase the AGN lightcurve exhibits a power-law like decay,
either due to the decrease in the fueling rate due to evolution of a
feedback-driven blast wave \citep{stochacc} or due to viscous evolution of an
accretion disk resulting from a quenched fuel supply
\citep[e.g.,][]{yu05,king07}. If the AGN lightcurve is decaying during the
Type 1 quasar phase, then we would expect a broad range of Eddington ratios
with the number densities increasing monotonically toward lower values of $L
/ L_{Edd}$. This is consistent with our estimated BHERF, although we note
that predictions for this type of BHERF are not unique to self-regulated
black hole growth models.

Our estimated BHERF is consistent with the $z \sim 1.4$ BHERF
  derived by \citet{nobuta2012} from a much deeper sample of X-ray
  selected Type 1 quasars. In addition, our BHERF is qualitatively
  consistent with the results of \citet{aird2012} and
\citet{bongiorno2012}. These authors found that the distribution of
X-ray luminosity at fixed stellar mass exhibits a power law-like
increase toward fainter X-ray luminosities for Type 2
quasars (in the case of \citet{aird2012}) and both Type 1 and Type 2
quasars (in the case of \citet{bongiorno2012}). Because $M_{BH}$ is correlated with stellar mass
\citep[e.g.,][]{mag98,haring04}, the results of \citet{aird2012} and
\citet{bongiorno2012} imply that the number densities of active SMBHs
increase toward lower Eddington ratio. This is in agreement with our
estimated marginal distribution of Eddington ratio, i.e., the
BHERF. However, these authors also concluded that the power-law
distribution of X-ray luminosity at fixed stellar mass is the same for
all stellar mass bins. This is inconsistent with our
result that for some of the redshift bins the distribution of Eddington ratio changes with
$M_{BH}$. That being said, the relationship 
between $M_{BH}$ and stellar mass is complex and exhibits
statistical scatter \citet{graham2012}, making a quantitative
comparison of our results (which are with respect to $M_{BH}$) with those of \citet{aird2012} and
\citet{bongiorno2012} (which are respect to stellar mass)
difficult. Moreover, both the \citet{aird2012} and
\citet{bongiorno2012} samples are X-ray selected and contain Type 2
quasars, and thus may probe a different population of objects. Thus,
is it unclear how inconsistent our results are compared with
theirs. In addition, the broad Eddington ratio distributions that 
we find at $z > 3$ may be inconsistent with small scatter in the
$L$--$M_{h}$ relation inferred from clustering measurements
\citep[e.g.,][]{white2008,shankar2010,shen2010,bonoli2010}. It is
unclear how to reconcile these results, especially if $M_{BH}$ and
$M_h$ are strongly correlated.

In addition, we have found that the distribution of $L / L_{Edd}$
is either approximately independent of $M_{BH}$ or shifts to larger
values with increasing $M_{BH}$, depending on the redshift bin, in contrast to the
results from continuity equation techniques \citep{shankar2012} or
from theoretical models \citep[e.g.,][]{granato2004,lapi2006}. Part of
this discrepancy may be that we only analyze the joint distribution of
$M_{BH}$ and $L / L_{Edd}$ for Type 1 quasars, while other techniques
tend to focus on the entire AGN population. This behavior could
represent a real effect, possibly due to evolution in the Type 1 quasar
lightcurve caused by, for example, a shallower decay in accretion rate
at $1 \lesssim z \lesssim 3$ during the blow-out phase for more
massive SMBHs. However, as discussed in \S~\ref{s-systematics} there
are several systematics that complicate analysis of the joint
distribution of Type 1 quasars in the mass luminosity plain, including
a possible dependence of the bolometeric correction on
$M_{BH}$ or $L / L_{Edd}$ and unknown systematic errors in the virial
mass estimates for \MgII\ and \CIV. In addition, the continuity
equation methods are also potentially affected by systematics as
discussed in \citet{bhmf_review}. Considering this, it is unclear if
the dependence of the Eddington ratio distribution on $M_{BH}$ that we
find is real, and these discrepancies highlight the need for further
improvement in our understanding of the virial mass estimates,
bolometeric corrections, and extensions of AGN demographical studies
to deeper surveys.

The downsizing in $L / L_{Edd}$ has a somewhat different form than that of
$M_{BH}$. Unlike the number densities of $M_{BH}$, the number densities in
bins of constant $L / L_{Edd}$ do not show any evolution at $z < 2$. Instead,
we observed `downsizing' in $L / L_{Edd}$ in the sense that Type 1 quasars
radiating at $L / L_{Edd} \lesssim 0.1$ are significantly more rare at $z
\sim 4$ compared to $z \lesssim 2$, while the number densities of Type 1
quasars at $L / L_{Edd} \gtrsim 0.1$ are similar for $z \sim 2$ and $z \sim
4$. This may reflect a stronger contribution to the BHERF at $z
  \sim 2$ from Type
1 quasars that are further along the decaying part of the AGN
lightcurve, and thus at a lower $L / L_{Edd}$, possibly due to a more
prolonged post-peak phase. In addition, this change in the $L /
L_{Edd}$ distribution at $z \lesssim 3$ may reflect a stronger
contribution of Type 1 quasars that are fueled through more internal
processes and experiencing weaker AGN activity. However, because our sample
becomes significantly incomplete below $L / L_{Edd} \lesssim 0.07$ it is
unclear how significant this trend is.

Based on our derived BHERF, we conclude that those Type 1 quasars that
do radiate near the Eddington limit are extremely rare, suggesting that if
Type 1 quasars do violate the Eddington limit they do so only for a very
brief period of time. In such super-critical accretion flows the
luminosity depends logarithmically on the accretion rate, so in
principle these Type 1 quasars that do radiate at $L / L_{Edd} > 1$ could be accreting at a significantly
higher rate relative to Eddington, i.e., $\dot{M} / \dot{M}_{Edd} \gg 1$.

\subsection{Comparison with Models for SMBH Growth}

Direct comparison of our results with models of SMBH fueling and growth is
difficult as most models do not present a quantitative prediction of the BHMF
and BHERF specifically for Type 1 quasars. Instead, many modelers predict the BHMF and
BHERF for all SMBHs, or for active SMBHs. Generically, many models predict a
BHMF which increases toward lower mass down to $M_{BH} \lesssim 3 \times 10^8
M_{\odot}$, which is in agreement with our estimated BHMF. In addition, many
models predict downsizing in SMBH growth and AGN activity, in agreement with
our empirical results. \citet{natarajan2012} compared BHMFs for Type 1 quasars with models for
merger driven SMBH growth that assumed that either SMBH seeds are the
remnants of Pop-III stars or were formed through direct collapse of
pre-galactic disks \citep{lodato07}. They compared their model unobscured
quasar BHMF to those estimated by K10, and in general they found that the
Pop-III seeding model underpredicted the BHMF at $M_{BH} \gtrsim 10^9
M_{\odot}$. Their direct collapse seeding model provided a better fit at $z <
4$ although it overpredicts the number densities of the most massive SMBHs in
Type 1 quasars at $z \lesssim 2$; however, the authors argue that the
overprediction is not a problem as it results from the fact that they did not
implement depletion of the available gas to grow the SMBH at lower $z$. None
of their models were able to match the K10 BHMF at $z = 4.25$. Because the
K10 BHMF was derived from the SDSS DR3, which is a subset of our sample, it
is not greatly different from the BHMF that we derive here. As such, the
conclusions reached by \citet{natarajan2012} would be unchanged using our
newer BHMF, although we note that comparison with our new BHMF implies that
all of their models underpredict our derived number densities of SMBHs in
Type 1 quasars with $M_{BH} \sim 3 \times 10^8 M_{\odot}$ at $z = 1.25$.

The most massive SMBHs in the observable Universe implied by our
BHMF have masses $10^{10} M_{\odot}$--$10^{11} M_{\odot}$, consistent with
earlier results obtained by K10. However, as this value is strongly
dependent on our prior for the extreme high mass tail of the BHMF, a
more realistic constraint is $M_{BH}^{\rm max} \gtrsim 10^{10}
M_{\odot}$. Recently \citet{mcconnell2011} detected two
SMBHs in the local universe with $M_{BH} \sim 10^{10} M_{\odot}$ determined
from stellar dynamical modeling. Considering the different volumes probed by
the SDSS DR7 and their search, their results are consistent with the
limits on the maximum mass implied by our BHMF. In addition, the lower
limits on the maximum masses that we infer are
consistent with values predicted from models that assume the SMBH's growth is
self-regulated \citep{natarajan2009}. Simulations that follow the growth of
SMBHs in bright $z \sim 6$ quasars have also been able to grow SMBHs to
$M_{BH} \sim 2 \times 10^{10} M_{\odot}$ by $z \sim 2$ \citep{sijacki2009}.

\subsection{Sources of Systematic Error}
\label{s-systematics}

While we have obtained a number of interesting results, there are a few
caveats regarding our approach that must be kept in mind, and we conclude
this section with a discussion of them. There are three significant potential
sources of systematic error in our approach, which we have touched on
earlier: incorrect specification of the virial mass error distribution and
bias, errors in the SDSS selection function, and errors in the bolometeric
correction. In this work we have used a simple constant bolometeric
correction to the 2500\AA\ luminosity, and, strictly speaking, our derived
BHERF should be viewed as the number density of Type 1 quasars radiating at a
given ratio of optical luminosity to Eddington, scaled upward by a constant
factor of five. Because the bolometeric correction likely depends on
both $M_{BH}$ and $L / L_{Edd}$ this will create systematic errors. It is
unclear how these systematics in the bolometeric correction affect our
results on the BHERF. In principle it is possible to incorporate a variable
bolometeric correction into our Bayesian model, but unfortunately there is
considerable uncertainty in the distribution of bolometeric corrections as a
function of $M_{BH}$ and $L / L_{Edd}$.

As discussed in \S~\ref{s-mbh_ndens} and \S~\ref{s-erat_ndens} systematic
errors on the selection function can have a significant effect on our results
in highly incomplete regions of parameter space. In general we have limited
our analysis to masses or Eddington ratios in which we are not highly
incomplete (i.e., completeness $\gtrsim 10\%$), so we do not expect
small errors in the selection function to have a significant effect on our
results. However, this is not true in the redshift bins corresponding to $z =
(2.15, 2.65, 3.20)$. In these bins the SDSS selection algorithm has
difficulty distinguishing quasars from stars, and as a result these bins are highly
incomplete. Moreover, the distribution of quasar colors obtained from the
simulations used to estimate the selection function do not perfectly match
the observed distributions, and the selection algorithm is color-dependent at
these redshifts. We do not include a color-dependence in the
selection function, and, because the quasar SED depends on $M_{BH}$ and $L /
L_{Edd}$, this likely introduces systematic errors into our incompleteness
correction. Because of this the BHMF and BHERF derived at these redshifts
should be interpreted with caution.

The most important possible source of systematic error in our approach arises
from incorrectly specifying the error distribution of the virial mass
estimates. This is particularly a concern for \MgII\ and \CIV, which are not
as well studied with respect to the reverberation mapping database as \hbeta.
In addition, \CIV\ emission is thought to at least partly arise from
an accretion disk wind and may contain a non-virial component to the line
width \citep[e.g.,][]{shen08,richards2011}, although it is currently unclear
if this creates a significant bias in the \CIV-based mass estimates. We have
tried to incorporate some systematic errors resulting from a bias in the
virial mass estimates to the extant that it can be modeled as having a simple
luminosity-dependence. There were no significant differences in our main
scientific conclusions when we included a simple luminosity-dependent bias.
In addition, we have tried to mitigate the affects of some objects having
large systematic errors in the mass estimates through our use of a Student's
$t$-distribution as a model for the measurement errors on the mass estimates,
which downweights any outliers.

In general, the number densities of the virial mass estimates obtained
before correcting for incompleteness do not show discontinuities across redshift
bins when switching emission lines (Paper I). However, there are a few
discontinuities in the derived mass and Eddington ratio functions when
switching between emission lines, suggesting that small systematics between
the virial mass estimates of different emission lines may be manifested more
strongly in our derived BHMF and BHERF. These include a discontinuity in the
normalization of the BHMF when going from \hbeta\ to \MgII\ (Figure
\ref{f-bhmf_evol}), a discontinuity in the BHERF when going from \MgII\ to
\CIV\ (Figure \ref{f-bherf_evol}), and differences in the distribution of $L / L_{Edd}$ at fixed
$M_{BH}$ inferred from the three emission lines (Figure \ref{f-cprob_erat}).
While it may be tempting to conclude from this that one or more of the
emission lines do not give consistent results, this is not necessarily the
case. As discussed in \S~\ref{s-mbh_ndens}, the increase in number densities
in the redshift bins using \hbeta\ is likely at least in part caused by a
host galaxy contribution to the nuclear emission which is not accounted for
in the selection function, creating an excess of AGN in these redshift bins
as their nuclear emission gets boosted above the flux limit. The
discontinuity in the number densities across $z \sim 2$, corresponding to the
shift from \MgII\ to \CIV, also corresponds to a transition in the selection
function to redshifts where quasar colors are similar to star colors. As
discussed above and in \S~\ref{s-erat_ndens} this can lead to significant
systematic error in the incompleteness correction; indeed, these redshift
bins have the largest statistical errors as well.

The only systematic differences between the emission lines in the inferred
distribution which cannot be explained as systematics from incorrect
incompleteness correction is the anomalous behavior of the Eddington
ratio distribution at fixed $M_{BH}$. The distribution of $L /
L_{Edd}$ is independent of $M_{BH}$ for the two bins employing \hbeta,
shifts toward larger values of $L / L_{Edd}$ with increasing $M_{BH}$ for the redshift bins employing
\MgII, and then shifts to being independent of $M_{BH}$ again over the
redshift bins employing \CIV. Although it has been argued that \CIV\ is less
reliable than \MgII\ or \hbeta\ \citep[e.g.,][]{baskin2005,shen08,shen_liu2012}, in this case it is
\MgII\ which gives the discrepant result. It is unclear why this is the case,
and it may be that this observed trend represents real evolution in the joint
distribution of $M_{BH}$ and $L / L_{Edd}$. However, we consider it likely
that unidentified systematics are at least in part driving this trend.
Because of this, our current understanding of virial mass estimates may not
provide enough accuracy for inferring how percentiles of the Type 1 quasar
joint distribution vary in the mass-luminosity plane.

\section{SUMMARY}
\label{s-summary}

In this work we have employed a Bayesian analysis method to derive the black
hole mass and Eddington ratio functions for broad line AGN using a
uniformly-selected sample from the SDSS DR7. We used more flexible models than those in Paper I to test the robustness of the key conclusions of Paper I, and found that the main results in Paper I remain valid, although the constraints are weakened due to the more flexible models used in the current work. Our conclusions are summarized
as follows:
\begin{itemize}
\item The SDSS is $\lesssim 10\%$ complete at $M_{BH} \lesssim 3 \times
    10^8 M_{\odot}$ or $L / L_{Edd} \lesssim 0.07$, with some variation
    with redshift. Decreasing the magnitude limits to $i \sim 24$,
    similar to that of the \emph{COSMOS} spectroscopic surveys or the
    \emph{Pan-STARRS} medium-deep fields, reduces the mass and Eddington
    ratio incompleteness limits by about an order of magnitude.
\item There are a broad range of $M_{BH}$ and $L / L_{Edd}$ for Type 1
    quasars, and there is no evidence for a peak in the black hole mass
    or Eddington ratio functions down to the $10\%$ completeness limits
    of the SDSS sample. The number densities of Type 1 quasars
    continue to increase toward lower $M_{BH}$ and $L / L_{Edd}$ down to
    at least $M_{BH} \sim 5 \times 10^8 M_{\odot}$ and $L / L_{Edd} \sim
    0.07$, respectively.
\item Both the BHMF and BHERF show evidence for downsizing. Relative to
    the peak in the number densities at $z \sim 2$, the number densities
    of the most massive black holes in Type 1 quasars fall off fastest
    toward lower redshift, while the number densities of the less massive
    black holes fall off faster toward higher $z$. This implies that the
    most massive black holes were active first, and shut off their
    activity more rapidly after the peak.

  The number densities of Type 1 quasars are approximately constant at $z
  \lesssim 2$ for $L / L_{Edd} \gtrsim 0.05$; however, the number
  densities of Type 1 quasars radiating at $L / L_{Edd} \lesssim 0.1$
  fall off more rapidly toward higher redshift at $z \gtrsim 2$, possibly
  reflecting a smaller contribution from weaker AGN activity toward
  higher redshift.
\item 
  We constrain the maximum value of $L / L_{Edd}$ in a Type 1
  quasar to be $\sim 3$. Therefore, if quasars do violate the Eddington
  limit, they do so only mildly and for a short period of time.
\item
  At low $(z \lesssim 0.8)$ and high $(z \gtrsim 2.65)$ redshifts the Eddington ratio distribution at fixed
  $M_{BH}$ is approximately independent of $M_{BH}$. However,
  at redshifts $0.8 \lesssim z \lesssim 2.65$ $p(L / L_{Edd}|M_{BH})$
  shifts toward higher values of $L / L_{Edd}$ from
  $M_{BH} = 5 \times 10^8 M_{\odot}$ to $M_{BH} = 5 \times 10^9 M_{\odot}$. This therefore implies that at intermediate redshifts the
  shape of the Eddington ratio distribution changes such that the high
  $L / L_{Edd}$ tail becomes more dominant at higher $M_{BH}$. At low
  and high $z$ the shape of the high $L / L_{Edd}$ tail of the Eddington
  ratio distribution is independent of $M_{BH}$. The redshift
  dependence of this trend is unexpected and may be due to unidentified
  systematics among the emission lines used to calculate the FWHM-based
  virial mass estimates.
\item
  We do not find statistical evidence for the so-called
  ``sub-Eddington boundary'' in the quasar mass-luminosity plane
  claimed by \citet{steinhardt2010a}. The appearance of such a boundary
  in the ``observed'' distribution is caused by selection effects and
  errors in the virial BH mass estimates (Paper I). This reinforces our
  early conclusions in Paper I that one should not interpret the
  observed distribution directly.
\item Assuming a radiative efficiency of $\epsilon_r = 0.1$ and a seed
    black hole mass of $M_{BH} = 10^6 M_{\odot}$, the growth times for
    SMBHs in Type 1 quasars having $M_{BH} \gtrsim 5 \times 10^8
    M_{\odot}$ are comparable to or longer than the age of the universe
    at $z \gtrsim 1.8$. Here, the growth times were calculated assuming
    that SMBHs in a given mass bin accrete at a time-averaged rate that
    is equal to the mean Eddington ratio in that mass bin. These large
    growth times imply that prior to us observing them as Type 1 quasars,
    these SMBHs experienced a stage of accelerated growth (i.e.,
    with higher Eddington ratios).
\item Comparison of the $M_{BH}$ completeness limits of our sample with
    the local mass function of all SMBHs implies that our sample is
    dominated by SMBHs representing the high mass end of the BHMF, which
    reside in what are locally early type galaxies. This conclusion in
    combination with our results on SMBH growth times, is consistent with
    models by which SMBHs experience a massive fueling event which
    initiates obscured growth. The black hole's growth is self-regulated,
    persisting until either feedback energy unbinds the obscuring gas or
    all of the gas is consumed from star-formation, briefly revealing the
    massive black hole as a Type 1 quasar with a decaying lightcurve. This same fueling event leaves behind
    a spheroid, placing the SMBH on the $M_{BH}$--$\sigma_*$
    relationship. Because the SMBHs in our sample represent the high mass
    end of BHMF, this process may only be common among this mass range.
    In addition, the long growth times of $z \lesssim 0.8$ Type 1
    quasars with massive BHs and low Eddington ratios likely
    represent weaker AGN activity reignited by mass loss from evolved
    stellar populations.
\end{itemize}

The combination of our large uniformly-selected sample with our powerful
Bayesian method represents an important contribution to AGN demographic
studies, and we have used our sample and method in a two-paper series to
obtain constraints on the optical luminosity function, black hole mass
function, and black hole
Eddington ratio functions of Type 1 quasars. In addition, we have used our
sample and method to place constraints on the distribution and biases of the
virial mass estimates. The results and methods presented in Paper I and in this paper
represent important empirical tools for understanding black hole growth, for
comparison to theoretical models, and for planning future surveys. In many
ways systematic AGN demographic studies with respect to $M_{BH}$ and $L /
L_{Edd}$ are just beginning. Further improvement will result from using
future reverberation mapping campaigns to refine our understanding of virial
mass estimators, as well as applying methods similar to our Bayesian technique to current and
future deeper surveys with possibly multiwavelength data.

\acknowledgements

We would like to thank Desika Narayanan, Marta Volonteri, and Benny
Trakhtenbrot for helpful discussions, and Marianne Vestergaard, Tommaso Treu, Priya
Natarajan, Marta Volonteri, Andreas Schulze, and Alister Graham for
helpful comments on drafts of our manuscript. We would also like to thank an anonymous referee for
comments that helped improve our clarity, focus, and discussion. BCK
acknowledges support from the Southern California Center for Galaxy 
Evolution, a multi-campus research program funded by the University of
California Office of Research. YS acknowledges support from the
Smithsonian Astrophysical Observatory (SAO) through a Clay Postdoctoral
Fellowship.

Funding for the SDSS and SDSS-II has been provided by the Alfred P. Sloan
Foundation, the Participating Institutions, the National Science Foundation,
the U.S. Department of Energy, the National Aeronautics and Space
Administration, the Japanese Monbukagakusho, the Max Planck Society, and the
Higher Education Funding Council for England. The SDSS Web Site is
http://www.sdss.org/.

Facilities: Sloan

\end{document}